\documentclass[12pt]{elsarticle}
\usepackage{latexsym}
\usepackage{booktabs}
\usepackage{geometry}
\usepackage{amsthm}
\usepackage{stmaryrd}
\usepackage{caption}
\usepackage[hyphens]{url}

\usepackage{setspace}
\usepackage{subfig}
\usepackage{color}
\usepackage[export]{adjustbox}
\usepackage{listings}
\usepackage{epstopdf} 
\usepackage{todonotes}
\usepackage{mathtools}
\usepackage[ruled,linesnumbered]{algorithm2e}

\usepackage{graphicx}
\usepackage{epstopdf}
\usepackage{multirow}
\usepackage{paralist}
\usepackage{amsmath,amsthm,amssymb}
\usepackage{setspace}
\usepackage{float}
\usepackage{morefloats}
\usepackage[]{algorithm2e}
\usepackage{mdwlist}
\usepackage{booktabs} % FvW: makes tables look nicer
\usepackage{amsfonts}
\usepackage{paralist}
\usepackage{soul}
\newcommand{\be}{\begin{equation}}
\newcommand{\ee}{\end{equation}}

\usepackage[final]{changes}
\newcommand{\RNum}[1]{\uppercase\expandafter{\romannumeral #1\relax}}
\theoremstyle{remark}

\usepackage{lineno,hyperref}
\modulolinenumbers[1]
\usepackage{geometry}
\usepackage{subfig}
\usepackage{color}
\usepackage[export]{adjustbox}
\usepackage{listings}
\usepackage{epstopdf} 
\usepackage{csquotes}
\newcommand{\highlighttext}[1]{#1}

\usepackage{enumitem}
\setlist[enumerate]{itemsep=0mm}
\setlist[1]{labelindent=0.5in,leftmargin=*}
\setlist[2]{labelindent=0in,leftmargin=*}

\usepackage{amsmath}
\usepackage{booktabs,amsfonts,dcolumn}
\makeatletter
\renewcommand{\fnum@figure}{\textbf{FIGURE~\thefigure} }
\renewcommand{\fnum@table}{\textbf{TABLE~\thetable} }
\makeatother

\usepackage{times}
\usepackage[T1]{fontenc}
\usepackage{textcomp}
\usepackage{todonotes}
\usepackage{multirow}
\usepackage{float}
\usepackage{lineno}

\usepackage{url}
\makeatletter
\def\url@leostyle{%
  \@ifundefined{selectfont}{\def\UrlFont{\sf}}{\def\UrlFont{\small\ttfamily}}}
\makeatother
\setcitestyle{round}

\begin{document}

	\begin{frontmatter}
        \title{Impact of Transportation Network Companies on Labor Supply and Wages for Taxi Drivers}
         \author[pu_address]{Lu Ling}
 		\ead{ling58@purdue.edu}
 		\author[ua_address]{Xinwu Qian}
 		\ead{xinwu.qian@ua.edu}
 		\author[pu_address]{Satish V. Ukkusuri}
 		\cortext[mycorrespondingauthor]{Satish V. Ukkusuri is the corresponding author.}
 		\ead{sukkusur@purdue.edu}
		\address[pu_address]{Lyles School of Civil Engineering, Purdue University, 
  West Lafayette, IN, 47907}
      \address[ua_address]{Department of Civil, Construction and Environmental Engineering, The University of Alabama, Tuscaloosa, AL, United States, 35487}
  
		\begin{abstract}
		    While the growth of TNCs took a substantial part of ridership and asset value away from the traditional taxi industry, existing taxi market policy regulations and planning models remain to be reexamined, which requires reliable estimates of the sensitivity of labor supply and income levels in the taxi industry. This study aims to investigate the impact of TNCs on the labor supply of the taxi industry, estimate the wage elasticity, and understand the changes in taxi drivers' work preferences. We introduce the wage decomposition method to quantify the effects of  TNC trips on taxi drivers' work hours over time, based on taxi and TNC trip record data from 2013 to 2018 in New York City. The data are analyzed to evaluate the changes in overall market performances and taxi drivers' work behavior through statistical analyses, and our results show that the increase in TNC trips not only decreases the income level for taxi drivers but also discourages their willingness to work. We find that 1\% increase in TNC trips leads to 0.28\% reduction of monthly revenue of the yellow taxi industry and 0.68\% decrease in monthly revenue of the green taxi industry in recent years. More importantly, we report that work behavior of taxi drivers shifts from the widely accepted neoclassical standard behavior to the reference-dependent preference (RDP) behavior, which signifies a persistent trend of loss in labor supply for the taxi market and hints at the collapse of taxi industry if the growth of TNCs continues. In addition, we observe that yellow and green taxi drivers present different work preferences over time. Consistently increasing RDP behavior is found among the yellow taxi drivers. Green taxi drivers were initially revenue maximizers but later turned into income targeting strategy.
		\end{abstract}
		
		
    \end{frontmatter}

	% \linenumbers
	% \pagewiselinenumbers % comment out for final manuscript; comment out if no line numbers on title page
	\thispagestyle{empty}

\section{Introduction}
\begin{figure}[h!]
\centering
\includegraphics[width=120mm]{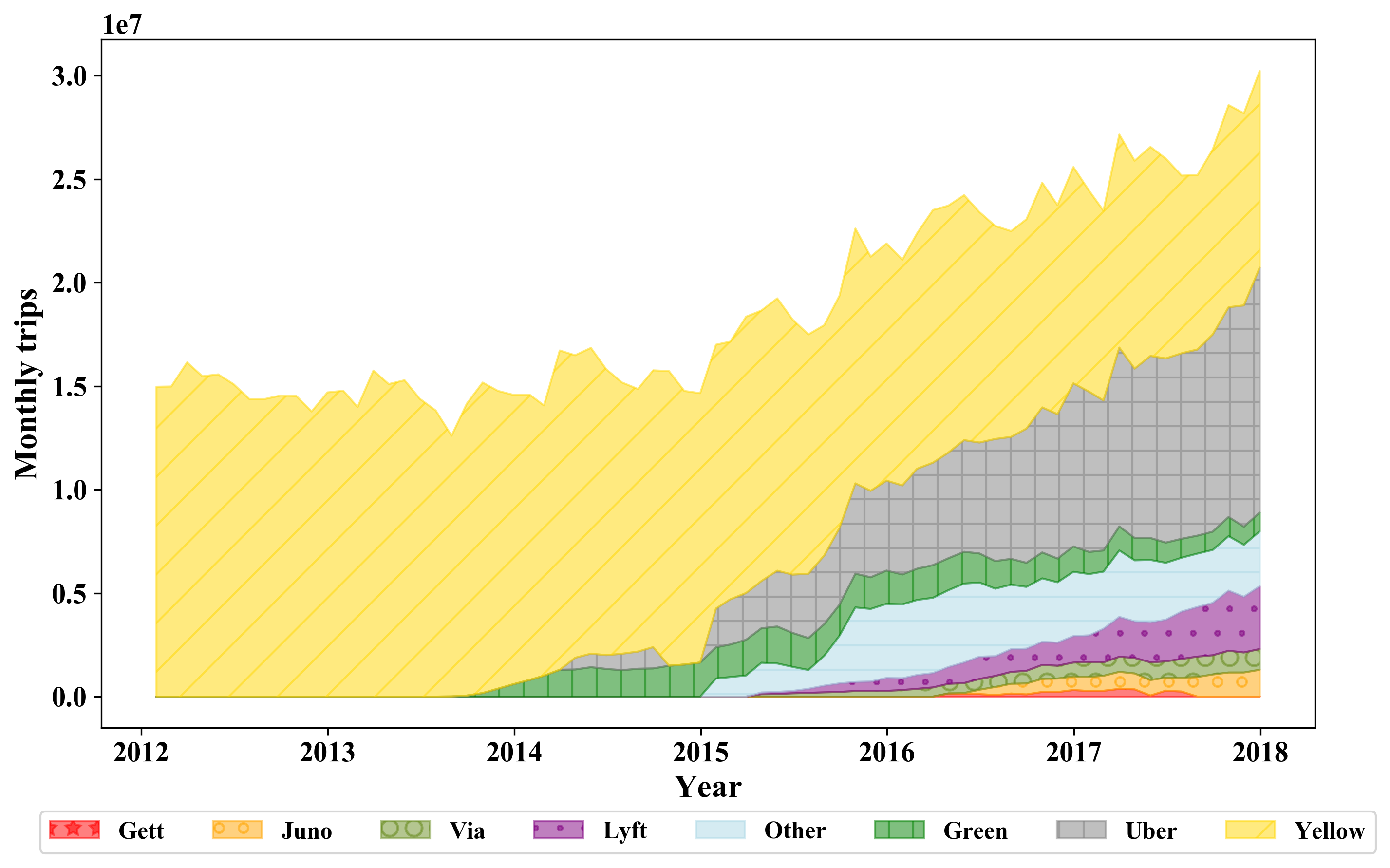}
\caption{Changes of monthly trips for major TNCs and taxis in NYC (2012-2017)}
\label{fig1}
\end{figure}

\begin{figure}[h!]
\centering
\includegraphics[width=120mm]{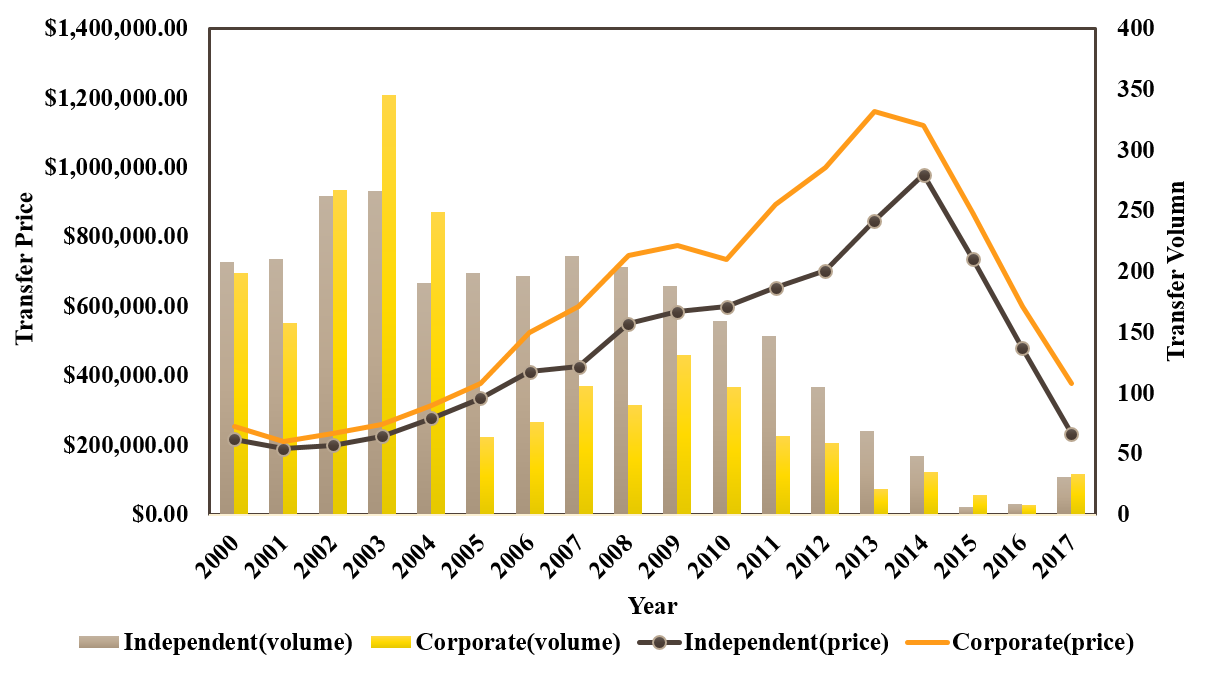}
\caption{Changes of medallion price and number of transferred medallions (2000-2017)}
\label{fig5}
\end{figure}

The recent rise of Transportation Network Companies (TNCs), such as Uber and Lyft, offers mobility-on-demand services to the general public, which has greatly changed the way people travel and has resulted in numerous externalities.  \highlighttext{Aside from the widely perceived issues such as heavier road congestion and safety concerns~\cite{hall2018uber,qian2020impact,lai2020evaluating,lai2020resilient,ukkusuri2020performance,ling2023influencing,ling2017reliable}, a recent study~\cite{Dan2018} revealed the significant loss for the taxi industry introduced by the involvement of TNCs' competitions, and the lack of regulations.The taxi industry suffered from losses in market share, ridership, labor supply, and asset values.} For instance, the leading TNCs in San Francisco had taken over almost $2/3$ of the original taxi market share between 2012 and 2014, which had caused the bankruptcy of the largest taxi operator in the city~\cite{timmurphy.org}. The official trip-level data released by New York City Taxi and Limousine Commission (NYCTLC) also disclosed that the taxi ridership is experiencing a steady decline since January 2009, as shown in Figure~\ref{fig1}. Along with the drop in taxi ridership, the value of NYC taxi medallion has dropped by 80\% since May 2013, which is a sharp contrast to the upward trend from 1975 to 2013 with a price of taxi medallion increased by 27 times~\cite{Barry2019} (see Figure~\ref{fig5}). As a consequence, the medallion transactions declined sharply after 2013~\cite{Dan2018}. Other taxi operators have reported 50\% idle rate of their medallions due to the lack of drivers~\footnote{Source: https://www1.nyc.gov/site/tlc/businesses/medallion-transfers.page}, creating the so-called taxi graveyard in the city~\cite{WF2015}. The significant loss of businesses in the traditional taxi market has put the taxi drivers in debt, resulted in casualties~\cite{Nikita2018}, and exacerbated the conflicts between taxi and TNC drivers. \highlighttext{In light of these issues, several regulation policies, such as the \enquote*{Tolling Program} in 2019 and the NYC 2020 \enquote*{No Action} baseline, are underway to control the overgrowth of TNCs and to promote a more sustainable ride-hailing market with the co-existence of TNCs and traditional taxi services.}

%NYC Department of Transportation (NYCDOT) has enacted a series policies to regulate the service of TNCs, including NYC 2020 'No Action' baseline} \footnote{Source: Improving Efficiency and Managing Growth in New York's For-Hire Vehicle Sector, made by New York City Taxi and Limousine Commission and Department of Transportation at June 2019. A set of future actions is mentioned in this report. In particular, the regulations about issuing new licenses for For-Hire-Vehicle require to define the cap of both FHV and taxi. Besides, the 'Tolling Program' has enacted in April 2019, where the FHV and taxi have different congestion charges. However, the effect of this action is waiting for evaluation}. Therefore, understanding the impacts of TNCs on the labor supply of taxi market is urgently needed and important, which provide underlying evidence for the policy formulation and implementation such as cap regulation and fee management for TNCs and taxi.

%\includegraphics[width=\linewidth]{fig/MEDAL.pdf}

%While much attention has been paid to the changes from the demand side of the taxi industry~\cite{contreras2017effects,cetin2013economic}, little effort is made in exploring the impact of TNCs on the labor supply of taxi market.

\highlighttext{Lying at the core of effective taxi market regulations is our understanding of the labor supply behavior, i.e., the number of active taxi drivers and taxi drivers' work hours.}
While much attention has been paid to the changes from the demand side of the taxi industry~\cite{contreras2017effects,cetin2013economic,ling2019forecasting}, few studies address the attention on the impact of TNCs on the labor supply of the taxi market. \highlighttext{Existing studies exaggerate the representative power of the historical data in predicting the future and modeling the labor supply behavior, such as dynamic pricing~\cite{qian2017time} and infrastructure planning~\cite{asamer2016optimizing}.} Other studies~\cite{frechette2019frictions,zha2016economic,qian2017taxi,farber2015you,yang1998network} introduce mathematical models or simulations to assist in framing taxi entry regulation and pricing policies based on the standard neoclassical model (NS), which assumes that taxi drivers follow revenue-maximizing behavior. They suggest that the labor supply function of taxi drivers is positively related to the trip revenue so that a transitory increase in the wage rate (wage per time unit) results in an increase in their work hours, and such behavior is supported by real-world observations when the taxi market was monopolistic~\cite{farber2008reference}. \highlighttext{These studies assume labor supply behavior being static, even though the competition from TNCs would potentially make their behavior rather dynamic. Nevertheless, due to the fact that taxi drivers are self-employed with freedoms in working hours management, evidence~\cite{crawford2011new} indicated that taxi drivers are likely to present reference-dependent-preference (RDP) behavior whenever their working environment changes. This evidence speculated that drivers' labor supply is negatively related to the wage rate as their wage approaches to a reference point. Considering the increasing number of TNC trips in recent years,} \highlighttext{\added{taxi drivers' labor supply behavior may have changed dramatically} and the NS behavior assumption based on the historical taxi record data may no longer be valid. As the executives decided not to cut the wage to a reference point during the Great Recession in order to prevent the discouragement of taxi drivers~\cite{eliaz2014reference,bewley2009wages}, we expect the history would repeat itself during the modern age. As Figure~\ref{fig5} indicates, however, the recent drastic drops in transfer price and the number of taxi medallion can discourage the willingness to work among taxi drivers. In this regard, the overall taxi market supply and individual work behavior may have changed due to the competition from TNCs. Meanwhile, drivers' labor supply behavior serves as a crucial input for modeling the competition between TNCs and the taxi industry, and our accurate interpretation of such behavior will directly contribute to effective regulations of the TNCs and taxi market. It is therefore of vital importance to investigate the labor supply behavior and the corresponding wage elasticity and explore how taxi drivers distribute their work hours under competition in the current taxi market.} \added{This motivates us to systematically quantify the changes of labor supply in the taxi market due to the impact of the TNCs, and we propose the following three research questions:}
\begin{itemize}
    \item How much do TNCs impact the overall labor supply and revenue of the taxi market?
    \item Do drivers decrease their expected wage along with the increasing number of TNC trips?
    \item Is the RDP behavior present among taxi drivers with the increasing number of TNC trips?
\end{itemize}

\added{To answer these research questions, we choose the taxi industry in NYC as a case study to understand the impacts of the TNC trips on the labor supply of the taxi market. For the first research question, we investigate how the competition from the TNCs, measured by the number of monthly TNC trips, affects the total work hours and revenue of both the yellow and green taxi markets. These analyses reveal what aspects of the taxi labor supply are significantly affected and how the effects change over time. While the cap for the NYC yellow taxi does not change, we observe a decline in the daily number of active drivers. This motivates us to further investigate the second and third questions, which are the effects of the growth of TNC trips on the taxi drivers' labor supply behavior.} \added{These two questions are examined by analyzing anticipated wage variation and wage elasticity using the wage decomposition method}. To support our analyses, we mine the NYC taxi and TNC trip record between January 2013, when TNCs were in their beginning stage, and December 2018, by the time that TNCs have accounted for over 50\% of the total mobility-on-demand market share (see Figure~\ref{fig1}). \highlighttext{\added{Our results show that 1\% increase in TNC trips in the current market will lead to 0.28\% decrease of monthly revenue in the yellow taxi market and 0.68\% decrease of monthly revenue in the green taxi market. Besides, 1\% increase in TNC trips in the current market will result in 0.29\% reduction of monthly work hours in the yellow taxi industry and 0.75\% reduction of monthly work hours in the green taxi industry.} Furthermore, the labor supply behavior of yellow and green taxis is different, where the unanticipated wage variation has increased by three times for yellow taxi drivers and by five times for the green taxi industry from January 2013 to December 2018}. \highlighttext{\added{This finding favors the co-existence of the NS and RDP behavior among taxi drivers, which supplements the existing literature for either supporting NS assumption or RDP assumption~\cite{farber2015you,crawford2011new}. Besides, it points out that the different labor supply behavior presents between yellow and green taxi drivers under the competition from TNCs, and the NS behavior is dominated at the TNCs' beginning stage. More importantly, it highlights the transition of individual behavior from NS to RDP due to the impact of TNCs, which is yet addressed by previous literature and indicates the potential collapse of the taxi industry if the extreme competition between taxi and TNCs continues.}} The insights have significant implications for taxi regulators and city policymakers in developing market planning models and framing regulatory policies on the cap and pricing management for taxis and for-hire vehicles (FHVs).

The rest of the study is organized as follows. Section 2 briefly reviews related studies. Section 3 outlines the data used in our study. \added{Section 4 introduces the methodology and corresponding hypotheses for the research questions. Section 5 presents the results of the analyses of the impact of TNCs on the overall taxi market. Section 6 presents the examination result of taxi drivers' labor supply behavior and wage elasticity under the presence and growth of TNCs. Section 7 summarizes the significant findings with concluding remarks.}
\section{Literature}
Though few studies have investigated the impact of TNCs on the labor supply of the taxi industry, there exists a broad literature on the discussion of the labor supply behavior of taxi drivers. 

From the macro-management perspective, researchers~\cite{contreras2017effects,cetin2013economic,ling2018analyzing} primarily focused on productivity improvements and taxi medallion utilization analyses. Meanwhile, there is an ongoing debate on the underlying behavioral assumptions for modeling the taxi driver objectives in the research community~\cite{douglas1972price,beesley1983information,yang2010nonlinear,yang2005regulating}. In an early work, Camerer et al.~\cite{camerer1997labor} presented a regression of log-transformation of drivers' daily work hours on log-transformation of average hourly earnings and characterized taxi drivers as having RDP behavior with a daily income target and quit driving once the target is met. Then, K\H{o}szegi and Rabin~\cite{kHoszegi2006model} proposed a theory of preferred personal equilibrium addressing that the anticipated wage will impact drivers' stopping decision (the drivers decide whether or not to look for an additional trip after the current trip). Besides, they clarified the anticipated wage is not status quo based but rational-expectation based~\cite{KszegiUtility}. Based on K\H{o}szegi and Rabin's theory, \highlighttext{Crawford and Meng~\cite{crawford2011new}} discussed a simplified point-based income target as well as target work hours utility function to model drivers' stopping behavior. In the study, they pointed out that taxi drivers' stopping decision is significantly related to hours but income as mentioned by Farber~~\cite{farber2005tomorrow}. In this direction, studies have also found that the labor supply curve of taxi drivers appears to slope downward~\cite{doran2014long,crawford2011new}, and the consensus of these works is that taxi drivers are targeting earners. For the NS model supporters, Solow~\cite{solow1956contribution} first proposed a positive relationship between labor supply and aggregated income. Later, Farber~\cite{farber2005tomorrow,farber2008reference,farber2015you} initially presented evidence for the existence of reference-dependence, but lately revised his study by analyzing the taxi supply pattern in weather-related conditions (e.g., rain and snow). And he deduced that taxi drivers are utility maximizers where drivers work more hours when the wage is high and work fewer hours when the wage is low. \highlighttext{Besides, Crawford and Meng~\cite{crawford2011new}} agreed that changes in the anticipated transitory wage would also be neoclassical. Ashenfelter et al.~\cite{ashenfelter2010shred} found an elasticity of -0.2 in response to the fixed fare change on the labor supply of NYC taxi drivers. Thakral et al.~\cite{thakral2017daily} speculated that more recent income has a stronger impact on drivers ending their shifts rather than income earned earlier. Buchholz et al.~\cite{buchholz2016semiparametric} presented that income-targeting behavior is associated with frequent shorter shifts, and the NS assumption is more suitable for explaining longer shifts. Recently, Frechette~\cite{frechette2019frictions} presented evidence of the reduction of market density when street-hailing and TNCs co-exist in the market based on the NS assumption.

These studies have provided valuable insights into patterns and mechanisms of labor supply in the taxi market. However, the labor supply analyses in these studies assume that the taxi market is still monopolistic, and the labor supply behavior of taxi drivers remains as the stationary NS assumption. However, the current taxi industry has undergone tremendous changes due to the competition from TNCs. And there is an emerging need to examine how the labor supply of the taxi market is affected and\added{ how taxi drivers' behavior changes} due to the growth of TNCs. In this study, we apply the datasets for both TNCs and taxis from January 2013 to December 2018 in NYC to address these issues. The taxi data depict \added{how the labor supply of taxi market changes and drivers' labor supply behavior varies over time.} The TNCs data are then used to mine the most affected aspects of taxi labor supply due to the rise of TNC trips. 

\section{Data}
The datasets used in this study are collected by NYCTLC\footnote{In 2009, NYCTLC initiated the Taxi Passenger Enhancement Project, which mandated the use of upgraded metering and information technology in all New York medallion cabs.}, which include both trip-level and aggregated shift-level information. The trip-level data contain the daily average yellow/green taxi trips, average trip duration, vendor ID, pickup and drop-off timestamps and locations (zonal level), trip distance, number of passengers, trip fare, tip, extra surcharges\footnote{These are miscellaneous extras and surcharges, and currently include  \$0.50 and \$1 rush hour and overnight charges.} in rush hour or overnight, improvement surcharge \footnote{The
improvement surcharge began being levied in 2015, and \$0.30 improvement surcharge assessed trips at the flag drop.}, tax, and tolls. Each valid trip is defined as at least one passenger in the car with the value of trip duration, trip distance, and trip fare to be a positive number.

\highlighttext{In the lack of individual shift-level data, we use the aggregated shift-level data in the study.} The aggregated shift-level data are monthly based and are processed from approximately 8.32 billion yellow taxi trips, 0.73 billion green taxi trips, and 6.55 billion TNC trips over six years (January 2013 to December 2018). The data are obtained at the average individual-level and market-level. \highlighttext{The average individual-level data cover daily average work hours per yellow/green taxi driver, daily average income per yellow/green driver, daily average yellow/green taxi medallions, monthly average active days per medallion. The average market-level data include the daily average fare of yellow/green taxis (include tips from credit card), monthly yellow/green taxi drivers, monthly yellow/green taxi medallions, total monthly work hours of all yellow/green taxi medallions and taxi drivers. We calculate the monthly income per driver by dividing the total fare per month by the monthly number of drivers. The data applied in our analysis can be seen in Table~\ref{tab:dataset}.}

\begin{table}[!h]
\centering
\caption{Data preparation}
\label{tab:dataset}
\begin{tabular}{p{3cm}p{6cm}p{6cm}}

\toprule
   & \multicolumn{1}{c}{Market-level} & \multicolumn{1}{c}{Average individual-level} \\
 \hline
 \multirow{12}{*}{Trip-level} & 
\begin{itemize}[leftmargin=*]
  \item {Daily average yellow/green taxi trips}
  \item Average trip duration
\end{itemize} & 
\begin{itemize}
  \item Vendor ID
  \item Pickup and drop-off timestamps and locations
  \item Distance
  \item Number of passengers
  \item Fare, tip, extra surcharge, improvement surcharge, tax, and toll
\end{itemize}
\\
 \multirow{14}{*}{Aggregated shift-level}& \begin{itemize}
  \item Daily average fare of yellow/green taxis
  \item Monthly number of yellow/green taxi drivers
  \item Monthly number of yellow/green taxi medallions
  \item Daily average yellow/green taxi medallions
  \item Total monthly work hours of all yellow/green taxi medallions and taxi drivers
\end{itemize}& \begin{itemize}
  \item Daily average work hours per yellow/green taxi driver
  \item Daily average income per yellow/green driver
  \item Monthly average active days per medallion
\end{itemize}
\\
 \toprule
\end{tabular}
\highlighttext{*Note: the individual shift-level data are not available.} 
\end{table}

%\multicolumn{2}{l}{\begin{tabular}[c]{@{}l@{}}Vendor ID, pickup and drop-off timestamps and locations, distance, number of passengers, fare, tip, \\ extra surcharges, improvement surcharge, tax, toll;\end{tabular}}
\section{Methodology}

\subsection{Hypotheses}
\begin{figure}[h!]
    \centering
    \includegraphics[width=170mm,trim={0.5cm 3.5cm 0cm 3.5cm},clip]{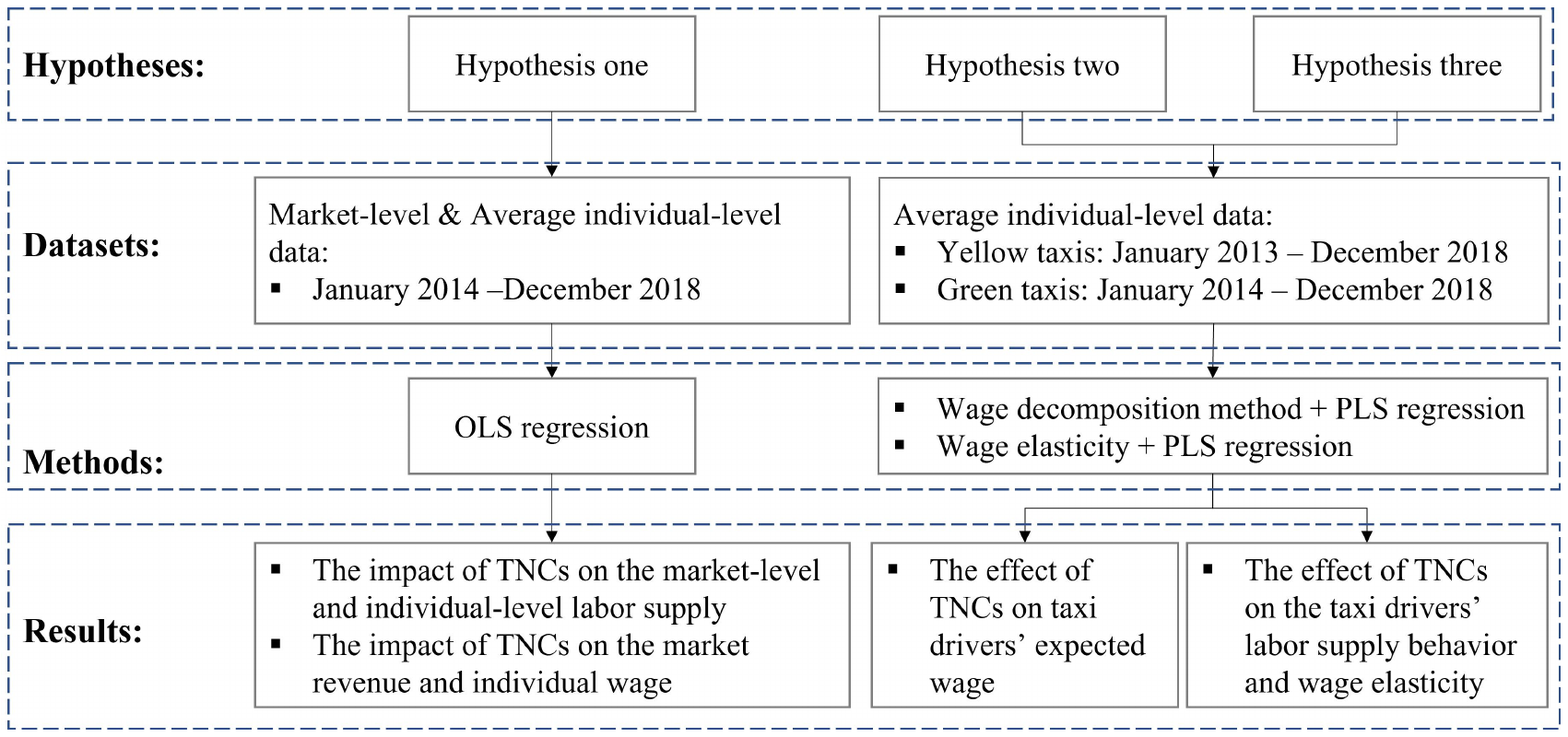}
    \caption{\highlighttext{The framework of the study}}
    \label{fig:framework}
\end{figure}

\added{The rapid growth of TNCs has significantly altered the landscape of the mobility-on-demand market and leads to a great loss in the taxi industry. To quantify how much do TNCs impact the overall labor supply and revenue of the taxi market, we propose the first hypothesis that \emph{\textbf{the rise of TNCs trips does not significantly impact the labor supply and revenue of the taxi market}}. The hypothesis is verified by a set of ordinary least square (OLS) regression and the fact that drivers' income has been significantly decreased while their work hours are barely affected contradicts the NS assumption. That is, drivers are not revenue optimizers, where their work hours are positively related to the wage rate. Because the same amount of work hours can not return the same amount of economic benefits as before. \highlighttext{Along this direction, we propose the second and third hypotheses to investigate drivers' labor supply behavior in the current market. Two hypotheses are tested based on the wage decomposition method via the partial least squares regression (PLS).} The second hypothesis is that \emph{\textbf{the increase of TNC trips does not decrease the taxi drivers' expected wage}}. The examination of the second hypothesis not only quantifies the effects of TNCs on drivers' expected wage but also provides a clue to RDP behavior among taxi drivers. \highlighttext{To inquiry whether the driver's labor supply behavior varies over time and how many drivers have RDP behavior in the current market, we propose the third hypothesis that \emph{\textbf{taxi drivers do not present RDP behavior}}.} The examination of the third hypothesis speculates taxi drivers' attitude toward the competitive ride-hailing market and implies the productivity of taxi drivers. } \highlighttext{The framework of the study is presented in Figure~\ref{fig:framework}}.

\subsection{Ordinary least square}
\highlighttext{To understand the aggregated labor supply changes under the impact of TNCs in the first research question, we adopt OLS regression as an unbiased estimation to infer the impact of TNC trips on the overall taxi market revenue and labor supply from January 2014 to December 2018. The estimated results can be found in section 5.1.} The natural log-transformation is conducted for TNC trips and the dependent variables so that the estimated coefficient can directly measure the elasticity of corresponding explanatory variables. The following model is proposed:
\begin{equation}\quad
ln(y_i)=a_1ln(x_{i,1})+\epsilon
\end{equation}
Where $y_i$ represents the $i_{th}$ observation of the dependent variable, $x_{i,1}$ is the $i_{th}$ observation of monthly TNC trips, and $\epsilon$ is the error term that is assumed to follow a normal distribution. The dependent variables are the collection of market-level and aggregated individual-level metrics for both yellow and green taxis. At the market-level, the dependent variables include the monthly taxi fare (revenue), monthly work hours for all taxi drivers, monthly taxi trips, monthly taxi drivers, monthly taxi medallions, average minutes per taxi trip, and average daily taxi medallions. At the average individual-level, the dependent variables involve monthly income per taxi driver, monthly work hours per taxi driver, monthly work hours per taxi medallion, average daily work hours per taxi driver, and average daily work hours per taxi medallion.
\highlighttext{\subsection{Reference-dependent preference}}
\highlighttext{RDP theory and NS theory are extensively used to describe the drivers' stopping decision, which describes the driver's decision whether or not to look for an additional trip after they finish current trip.} RDP theory has its root in prospect theory and indicates loss-aversion behavior, where the loss is considered more painful than the pleasure from the same amount of gain. \highlighttext{RDP suggests that drivers are rational at setting an income target and target work hours to decide whether or not to look for an additional trip after current trip. The income target in RDP describes that the driver works more hours in low wage rate days and works fewer hours in high wage rate days.} In contrast, the NS theory relies on the revenue-maximizing behavior and is a widely-used assumption in modeling drivers' behavior. It assumes a positive correlation between the drivers' work hours and wage rate and does not consider the targeting behavior.

\highlighttext{According to Crawford and Meng~\cite{crawford2011new}}, drivers' utility $V$ comprises the consumption utility $U_1 (I)+U_2 (H)$ and the gain-loss utility $R(I,H| I^r,H^r)$  with weights of $(1-\eta)$ and $\eta$, respectively $(0\leq \eta\leq 1)$. $I$ and $H$ are daily income and daily work hours. $I^r$ and $H^r$ denote the income target and the target work hours. The complete utility function of RDP can be written as:
\begin{equation}
V(I,H|I^r, H^r)=(1-\eta)(U_1(I)+U_2(H))+\eta R(I,H|I^r, H^r)
\end{equation}

The utility function of NS is:
\begin{equation}
V(I,H)=U_1 (I)+U_2 (H)
\end{equation}

According to K\H{o}szegi and Rabin's assumption~\cite{kHoszegi2006model}, consumption utility is additively separable across income and hours, with $U_1 (I)$ increasing in $I$ and $U_2 (H)$ decreasing in $H$, and both are concave. The coefficient of loss-aversion has a constant weight of $ \gamma(\gamma\geq 1)$ relative to gains. In addition, a linear relationship between expected income and hours is $I=w^e H$, where $w^e$ is the expected wage. In Farber's study~\cite{farber2015you}, the disutility of work hours can be expressed as:
\begin{equation}
    U_2 (H)=\frac{\theta}{(1+v)} H^{(1+v)}
\end{equation}
 where $\theta $ is the coefficient of disutility of work hours and $\frac{1}{v}$ is the wage elasticity of labor supply. Given any days, the gain-loss utility $R(I, H| I^r, H^r)$ have four scenarios as shown in Table~\ref{scenario}.
\begin{table}[!h]
    \centering
    \caption{Gain-loss utility $R(I,H|I^r,H^r)$ in different scenarios}
    \begin{tabular}{lcc}
    \toprule
         &  Income gain($I>I^r$) & Income loss($I< I^r$)\\
         \hline
        Hours gain($H<H^r$) &   Scenario 1 &  Scenario 2  \\
        Hours loss($H> H^r$) &    Scenario 4  &  Scenario 3  \\
        \bottomrule
    \end{tabular}
    \label{scenario}
\end{table}

The drivers’ optimal stopping decision is to maximize their utility $V$, which depends on the first-order condition of $V$ in each scenario. At the beginning of working day, if the driver works on a \enquote*{good day} with a realized wage rate higher than the expected wage ($w>w^e$), he passes through the income gain and hours gain domain (scenario 1), which is consistent with the NS model; if the driver works on a \enquote*{bad day} with a realized wage rate lower than the expected wage ($w<w^e$), the driver starts from the income loss and hour gain domain (scenario 2). Before reaching the target work hours $H<H^r$, the income loss motivates the driver to work. Thus, the driver's optimal work hours in scenario 2 is $ H=\left(\frac{(1-\eta+\eta\gamma)w}{\theta}\right)^{(1/v)}$ and wage elasticity of labor supply is $\frac{1}{v}$. Once the target work hours reaches ($H>H^r$) while the realized income is less than target income($I<I^r$), the labor supply curve of the RDP is the same as its NS curve, and the same as it is in the scenario 1 when the drivers are in \enquote*{good day}. Then, the optimal work hour is decided by $ H=\left(\frac{w}{\theta}\right)^{(1/v)}$ with wage elasticity being $\frac{1}{v}$. \highlighttext{When the income target is reached ($I=I^r$), and the realized hours are above the target work hours ($H>H^r$), the wage elasticity at this condition is -1. As K\H{o}szegi and Rabin mentioned, the RDP dominates when the elasticity is 0.} \highlighttext{When the income is sufficiently high enough ($I>I^r$) to reverse target work hours that reach the first time and work hours is greater than target work hours ($H>H^r$), which is scenario 4. The hour loss lowers the drivers' incentive to work in this condition. The optimal work hour is decided by:} $H=\left(\frac{w}{(1-\eta+\eta\gamma)\theta}\right)^{(1/v)}$ and the wage elasticity of labor supply is $\frac{1}{v}$.

 According to the preferred personal equilibrium theory~\cite{KszegiUtility}, the driver is more likely to work if there is a high probability that the future wage rate will increase. While the labor supply of taxi drivers would be negatively related to the wage rate when they face uncertainty~\cite{crawford2011new}. Thus, the gain-loss utility in RDP is only related to unanticipated transitory changes in wage. With the rise of TNCs trips, the driver will go through the scenarios in \enquote*{bad day} due to the competition, where the stopping decision for whether or not to look for an additional trip after the current trip is mainly determined by the income target~\cite{crawford2011new}. In our case, the basic idea is to investigate whether the taxi drivers decrease their expected wage under the growth of TNC trips based on the fact that taxi drivers' monthly work hours have barely affected while their monthly income decreases. Moreover, we are interested in how taxi drivers may respond to the current market, which is how many taxi drivers' behavior can be explained by the RDP and the NS. We follow the similar wage decomposition approach as applied in Farber's study~\cite{farber2015you} to investigate the change proportion of anticipated and unanticipated transitory wage variation.
 
 %This indicates that taxi drivers' labor supply in a time unit will be positively related to anticipated transitory wage changes. 

%Later, Farber~\cite{farber2015you} proposed the wage decomposition method to estimate the drivers' stopping decision based on the taxi dataset from 2009 to 2013. 

\subsection{Wage decomposition and wage elasticity}
The wage decomposition method investigates the second and third hypotheses and quantifies drivers' anticipated and unanticipated wage variations. As indicated by previous studies~\cite{kHoszegi2006model,farber2015you,crawford2011new}, the anticipated wage variation reflects the NS behavior and the RDP behavior is related to the proportion of unanticipated wage variation. Intuitively, the drivers' anticipated wage is expectation-based, which comes from the knowledge of the taxi market. The TNC trips, the improvement surcharge, the extra surcharges, and time variation will be the main aspects that impact drivers' judgments on the wage. Therefore, \highlighttext{the wage rate (wage per day) is decomposed by a two-stage process in the study}. In the first stage, we regress the natural log transformation of the average wage rate on the improvement surcharge indicator and extra surcharge indicator by PLS. These two surcharge indicators capture the fixed anticipated wage variation. Note that the residual of the first stage contains both anticipated and unanticipated transitory wage variation. In the second stage, we conduct PLS on the residual with the year indicators, month indicators, and the natural log transformation of monthly average TNC trips. The TNC trips over time affect the taxi drivers' judgment on temporal earning opportunities and expected wage. The variance captured by the second stage describes the anticipated transitory wage variation, and the remaining residual therefore represents the unanticipated transitory wage variation. Thus, the wage decomposition approach contains three parts: fixed anticipated wage variation, anticipated transitory wage variation, unanticipated transitory wage variation. The fixed anticipated wage and anticipated transitory wage are involved by the expected wage proposed in the second research question. The proportions of anticipated transitory wage variation and unanticipated transitory wage variation will be used to estimate the proportion of the RDP behavior and the NS behavior among taxi drivers, \highlighttext{and the estimates can be found in section 5.2. }

As for the data preparation, we take the data from January 2013 to December 2014 as the base year group for the yellow taxi estimation and January 2014 to December 2015 for green taxi estimation due to the limited number of TNC trips during this period. We further introduce two experiments to ensure a sufficient number of data samples and testify the consistency of our results. The dataset is grouped as 30 consecutive months for yellow taxis and as 24 consecutive months for green taxis. Two experiments follow the same wage decomposition approach as mentioned above. That is, the tested performance metric is the natural log transformation of drivers' average wage rate, and explanatory variables are the set of time indicators, natural log transformation of TNC trips, and the fare surcharge indicators. Thus, in experiment \uppercase\expandafter{\romannumeral1}, we have eight data groups for yellow taxis and six data groups for green taxis, which include: $ 01/13-06/15, 07/13-12/15, 01/14-06/16, 07/14-12/16, 01/15-06/17, 07/15-12/17, 01/16-06/18$, and $07/16-12/18$. In experiment \uppercase\expandafter{\romannumeral2}, we get nine data groups for yellow taxis and seven data groups for green taxis, which include: $01/13-12/14, 07/13-06/15,01/14-12/15, 07/14-06/16, 01/15-12/16, 07/15-06/17, 01/16-12/17, 07/16-06/18$, and $01/17-12/18$. Finally, the wage decomposition method is conducted on each of the groups.

Besides, we measure taxi drivers' wage elasticity as the percentage changes of monthly work hours to the changes in monthly wage. Therefore, the estimated coefficient describes the wage elasticity of the taxi labor supply. The datasets and experiments included in the analyses of wage elasticity are the same as the wage decomposition approach.

\subsection{Partial least squares}
\highlighttext{From real-world observations, drivers' wage rate expectation will be affected among a set of temporal indicators, fare surcharge indicators, and the TNC trips factor. In order to avoid overfitting issues and multicollinearity among variables, we apply the PLS regression to decompose their wage rate to gain an accurate estimate of the interactions between drivers' expected wage and the set of explanatory variables, and the results can be obtained in section 5.2. PLS regression was developed in the 1960s~\cite{wold2004partial} as an econometric method and is particularly suited for analyzing the system of equations that contain a vast number of indicator variables, in which the maximum likelihood covariance-based tools reach their limits.} The basic idea of PLS is to extract factors that account for as much variance in the response variables as possible~\cite{rashid2015methodological}. PLS considers both factors space and response space and seeks the dominant direction for the successive pairs between two spaces. Mathematically, it establishes a linear combination of the columns of $X$ and $Y$ such that their covariance is maximum~\cite{Haenlein2004AB}:
\begin{eqnarray}
X=TP^T+E\\
Y=UQ^T+F \\
u_i=b_it_i
\end{eqnarray}
where $P$ and $Q$ are termed as the loading matrices for $X$ and $Y$. $T$ and $U$ are the projections(scores) for $X$ and $Y$, respectively. $E$ and $F$ are the error terms and are assumed to be independent and identically distributed random normal variables. $u_i$ and $t_i$ represents the $i_{th}$ component of $U$ and $T$, and $b_i$ is the regression coefficient. In our model, $Y$ is the natural log-transformation of average daily income. $X$ is the set of variables, including the natural log-transformation of TNC trips, temporal indicators for years and months, the improvement surcharge indicator, and the surcharge indicator (during rush hours and overnight hours).

\section{RESULTS}
\subsection{TNCs impact on overall taxi market performances}
\begin{figure}[H]
     \centering
     \includegraphics[width=1\linewidth]{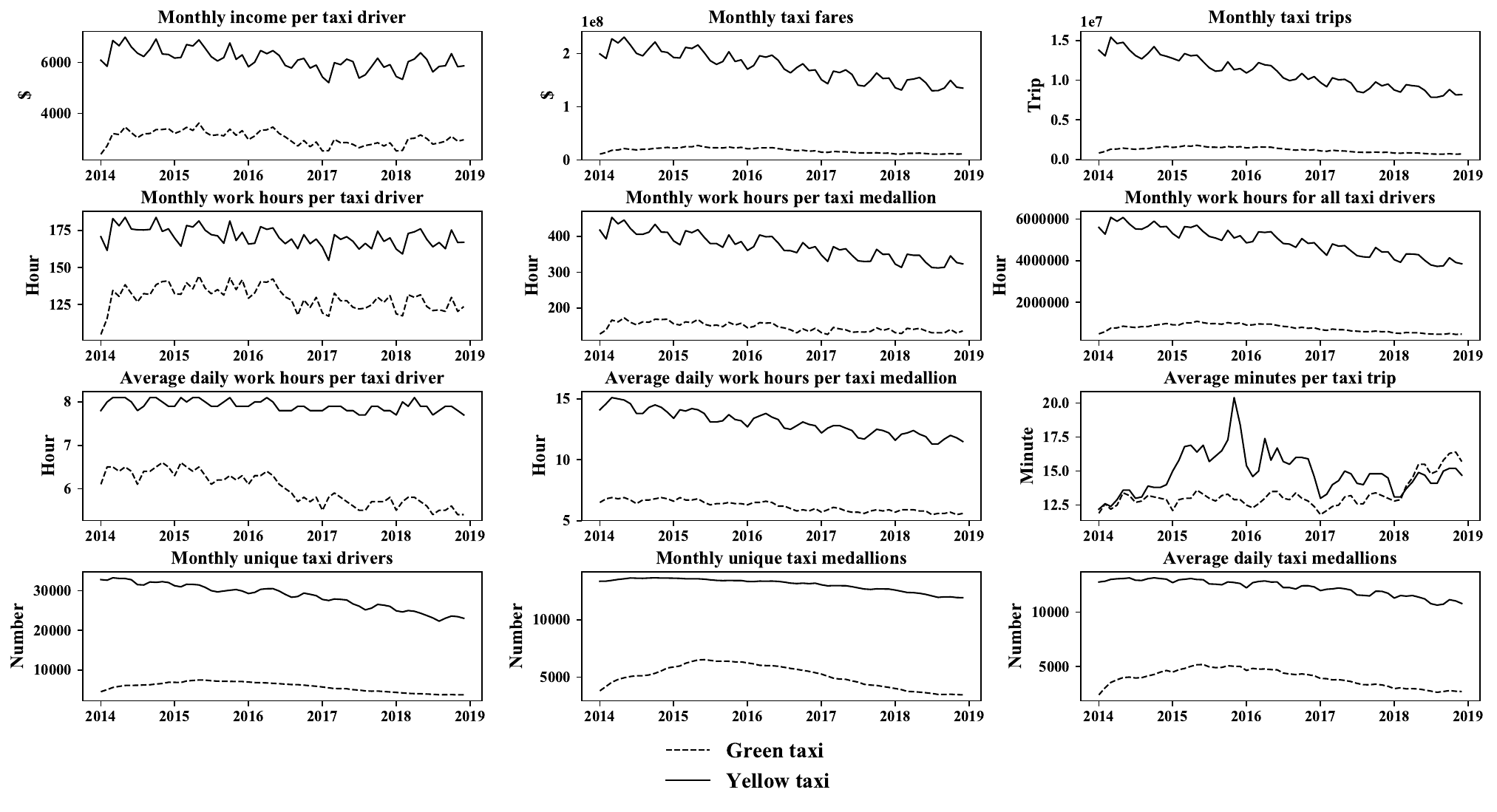}
     \caption{Overall taxi market performances from January 2014 to December 2018}\label{overall}
\end{figure}

\added{To explore the first research question, \emph{\textbf{how much do TNCs impact overall labor supply and revenue of the taxi market}}, we mine the overall taxi market performance metrics from January 2014 to December 2018 as shown in Figure~\ref{overall} and test the first hypothesis}. There is a consistent decrease in ridership, market revenue, and market total labor supply (work hours for both taxi drivers and taxi medallions) over time for yellow taxis. Besides, we observe a  substantial reduction in the monthly income per yellow taxi driver since the beginning of 2016, followed by a small income rebound in early 2018. Meanwhile, the monthly work hours per yellow taxi driver roughly remained the same over these years. In contrast to trends for the yellow taxi market, we find the increases in green taxi ridership before June 2015 and decreases after that time. As for the individual driver, the monthly income and monthly work hours are observed to first increase before June 2016 and then decrease rapidly. These results provide strong evidence that the nature of the taxi market has changed significantly in recent years. Meanwhile, the TNCs have seen their most rapid growth since early 2014. This overlapping of time consequently evokes a question of whether the rise of TNCs should account for the changes in taxi markets.

\highlighttext{\added{The hypothesis testing comprehensively measures the impacts of the increasing number of TNC trips on overall market performance metrics in the taxi industry.} The estimated results of the OLS regression in five years range are shown in Table~\ref{tab:olsresult_market} and Table~\ref{tab:olsresult_individual}. The OLS estimates take the monthly TNC trips as an indicator variable and regress on a set of taxi market responses, which include trips, fares, and labor supply from both the market-level and average individual-level for yellow and green taxis. }Several significant findings can be identified based on the results. First, the monthly yellow taxi trips are negatively related to the increase of the TNC trips (at 0.001 significance level). \highlighttext{Besides, 1\% increase in TNC trips would lead to 0.02\% reduction in yellow taxi trips (TNC trips increase 2.2\% from November to December 2018 and lead to about 46,000 yellow taxi trips reduced in a month). As for the market revenue and market total labor supply of the yellow taxis, both variables are observed to be significantly negative to the increase of TNC trips. We observe that 1\% increase in TNC trips would result in 0.02\% reduction in monthly fare and 0.01\% reduction in monthly work hours of the yellow taxi market (TNC trips increase 2.2\% from November to December at 2018 lead to about \$687,500 revenue reduction and about 11,550 work hours reduction for yellow taxi market in a month).} Moreover, the monthly yellow taxi drivers, medallions, as well as daily taxi medallions decreased rapidly. These findings suggest that the rise of TNCs takes a vast market share from the yellow taxi market. Contrary to the yellow taxis, we also note no significant impacts on the green taxi market in terms of ridership, market revenue, and total labor supply in five years.

\begin{table}[!h]
\centering
\caption{The impact of TNC trips on the market level performance of labor supply (OLS estimation)}
\label{tab:olsresult_market}
\begin{tabular}{lcccc}
\toprule
\multirow{2}{*}{Dependent variables}                                & \multicolumn{2}{c}{Yellow taxi} & \multicolumn{2}{c}{Green taxi} \\
\cmidrule(l){2-3} \cmidrule(l){4-5}
                                                                    & Coefficient     & Adj.$R^2$  & Coefficient     & 
                                                                    Adj.$R^2$  \\
                                                                    \hline
\multirow{2}{*}{Monthly taxi trips}        & -0.0219   & 0.338      & -0.0133   & 0.031      \\
                                                                    & (9.76E-06 ***)  &            & (0.095 .)
                                                &            \\
\multirow{2}{*}{Monthly taxi fares}    & -0.0166   & 0.245     & -0.0101    & 0.013      \\
                                                                    & (3.48E-05 ***)  &            & (0.185) &            \\
\multirow{2}{*}{Monthly work hours for all taxi drivers}                    & -0.0146   & 0.269      & -0.0078   & 0.004     \\
                                                                    & (1.32E-05 **)  &            & (0.269) &            \\

\multirow{2}{*}{Monthly taxi drivers}                 & -0.0123   & 0.276      & -0.0075    & 0.011      \\
                                                                    & (9.79E-06 ***)  &            & (0.206)  &            \\
\multirow{2}{*}{Monthly taxi medallions}               & -0.0035    & 0.156      & -0.0016   & -0.016      \\
                                                                    & (0.001 ***)  &            & (0.767) &            \\
\multirow{2}{*}{Average daily taxi medallions}      & -0.0056   & 0.198      & -6.27E-05   & -0.017     \\
                                                                    & (0.0002 ***)  &            & (0.991)  &           \\
                                                             \bottomrule
                                                                    
\end{tabular}\\
Note: the p-value of the estimated coefficient is shown in the bracket (. : $\leq$ 0.1; *: P $\leq$ 0.05; **: P $\leq$ 0.01; ***: P $\leq$ 0.001).
\end{table}

\begin{table}[!h]
\centering
\caption{The impact of TNC trips on the individual level performance of labor supply (OLS estimation)}
\label{tab:olsresult_individual}
\begin{tabular}{lcccc}
\toprule
\multirow{2}{*}{Dependent variables}                                & \multicolumn{2}{c}{Yellow taxi} & \multicolumn{2}{c}{Green taxi} \\
\cmidrule(l){2-3} \cmidrule(l){4-5}
                                                                    & Coefficient &    Adj.$R^2$  & Coefficient     & 
                                                                    Adj.$R^2$  \\
                                                                    \hline
\multirow{2}{*}{Monthly income per taxi driver} & -0.0044   & 0.09     & -0.0026  & 0.002      \\
                                                                    & (0.011 *)  &            & (0.290)   &            \\
\multirow{2}{*}{Monthly work hours per taxi driver}               & -0.0023   & 0.075     & -0.0004    & -0.016      \\
                                                                    & (0.020 *)  &            & (0.828)  &            \\
\multirow{2}{*}{Monthly work hours per taxi medallion}               & -0.011   & 0.300     & -0.0064    & 0.121      \\
                                                                    & (3.57E-06 ***)  &            & (0.004 **)  &            \\
                                                                    
\multirow{2}{*}{Average daily work hours per taxi medallion}         & -0.0091   & 0.300      &-0.0082    & -0.31     \\
                                                                    & (6.99E-07 ***)  &            & (2.23E-06 ***) &            \\
\multirow{2}{*}{Average daily work hours per taxi driver}      & -0.0009   & 0.065      & -0.0066   & 0.261     \\
                                                                    & (0.028 *)  &            & (1.82E-05 ***)  &           \\                                                                    
\multirow{2}{*}{Average minutes per taxi trip}                     & 0.0085   & 0.162     & 0.0043    &0.064      \\
                                                                    & (0.001 ***)  &            & (0.028 *)&            \\
                                                                     \bottomrule

\end{tabular}\\

Note: the p-value of the estimated coefficient is shown in the bracket (. : $\leq$ 0.1; *: P $\leq$ 0.05; **: P $\leq$ 0.01; ***: P $\leq$ 0.001).
\end{table}

From the individual perspective, the monthly income and work hours per yellow taxi driver are observed to decrease 0.04{\textperthousand}  and 0.02{\textperthousand} along with 1\% increase of the TNC trips. And from the five years range, the rise of TNC captures a small proportion of the variation of yellow taxi drivers' monthly income and work hours. Moreover, the impacts on the utilization (work hours) of per taxi medallion are much more significant (at 0.001 significance level) than it is on the monthly work hours per yellow taxi driver (at 0.05 significance level), and green taxis yield the same result. In combination with the observed drastic reduction in total labor supply of the yellow taxi market (0.01\% reduction along with
1\% increase of TNC trips), such evidence implies that there is only a minor change in the monthly work hours per individual yellow taxi driver (0.02{\textperthousand} reduction along with 1\% increase of TNC trips) while a significant reduction in the monthly driver (0.01\% reduction along with 1\% increase of TNC trips) is observed. The impacts on monthly labor supply and monthly income per driver are not statistically significant for green taxis. However, daily work hours per green taxi driver are found to decrease significantly by 0.07{\textperthousand} along with 1\% increase in monthly TNC trips. This suggests that individual drivers work more days in a month since there are no significant changes in the monthly work hours.

The above results are based on January 2014 and December 2018, where we have seen the most rapid growth of the TNC sector. \highlighttext{The well-known fact is that the taxi market has lost significant ridership, and we can also verify that the rise of TNCs has significantly decreased total market revenue and labor supply for yellow taxis from the data. Nevertheless, such impact is found to be non-significant for the green taxi market, which is a special class of taxi service for serving areas outside Manhattan. These observations lead to two important implications.} First, the taxi demand within Manhattan, which used to be served exclusively by yellow taxis, is close to its saturated level, and the TNCs, therefore, directly compete with yellow cabs for existing passengers. However, for areas outside Manhattan, we find empirically that the TNCs lead to induced ridership that is unsatisfied before and contribute to mitigating the under-supply issue in these areas. 
\begin{table}[!h]
\centering
\caption{Yellow taxi revenue and labor supply variation (OLS estimation)}
\label{tab:yellow taxi market time period variation}
\begin{tabular}{lcccc}
\toprule 

\begin{tabular}[c]{@{}c@{}}Month/Year\end{tabular} & \begin{tabular}[c]{@{}c@{}}Monthly fare\\all taxi drivers \end{tabular} & \begin{tabular}[c]{@{}c@{}}Monthly income\\per taxi driver\end{tabular} & \begin{tabular}[c]{@{}c@{}}Monthly work hours\\all taxi drivers\end{tabular} & \begin{tabular}[c]{@{}c@{}}Monthly work hours\\per taxi driver\end{tabular} \\
\hline

\multirow{2}{*}{01/13-06/15}& -0.0013& 0.0018& -0.0019 &0.0004\\
                            &(0.421)&(0.434)&(0.125)&(0.673)
\\
\multirow{2}{*}{07/13-12/15} & -0.003& 0.0006&-0.0035&0.0001 \\
                           & (0.078 .)&(0.608)&(0.01 *)&(0.867)
\\
\multirow{2}{*}{01/14-06/16} & -0.0042& -0.0003& -0.0045&-0.0005
\\
                           &(0.055 .) & (0.856) &
                           (0.01 *) &(0.597)\\
\multirow{2}{*}{07/14-12/16} & -0.0079 &-0.0032&-0.0073&-0.0025                                                                           \\
                            &(0.008 **)                                                      & (0.091 .)                                                         & (0.002 **)  &(0.036 *)                                                        \\

\multirow{2}{*}{01/15-06/17} & -0.1308&-0.0615&-0.0914&-0.0221                                                                         \\
                           &(1.033E-05 ***)                                                      & (0.002 **)                                                         & (3.26E-05 ***) &(0.066 . )                                                         \\

\multirow{2}{*}{07/15-12/17} & -0.1913&-0.0668&-0.1469&-0.0238                                                                         \\
                           &(6.77E-05 ***)                                                      & (0.02 *)                                                         & (4.36E-05 ***) &(0.261)                                                         \\
\multirow{2}{*}{01/16-06/18} & -0.2925                                                      & -0.0392                                                           & -0.2552  &  -0.0018                                                      \\
                           &(3.87E-06 ***)                                                      & (0.325)                                                         & (1.73E-07 ***) &(0.937)                                                         \\
\multirow{2}{*}{07/16-12/18} & -0.2785 &0.0398&-0.2896&0.0287                                                                  \\
                           &(2.08E-05 ***)                                                      & (0.323)                                                         & (1.94E-07 ***)  &(0.221)                                                        \\

                           \bottomrule
\end{tabular}

Note: the p-value of the estimation of monthly TNC trips is in the bracket(. : $\leq$ 0.1; *: P $\leq$ 0.05; **: P $\leq$ 0.01; ***: P $\leq$ 0.001).
\end{table}

\begin{table}[!h]
\centering
\caption{Green taxi revenue and labor supply variation (OLS estimation)}
\label{tab:green taxi market time period variation}
\begin{tabular}{lcccc}
\toprule
Month/Year          & \begin{tabular}[c]{@{}c@{}}Monthly fare\\ all taxi drivers \end{tabular} & \begin{tabular}[c]{@{}c@{}}Monthly income\\  per taxi driver\end{tabular} & \begin{tabular}[c]{@{}c@{}}Monthly work hours\\ all taxi drivers\end{tabular} &
\begin{tabular}[c]{@{}c@{}}Monthly work hours\\ per taxi driver\end{tabular} \\
\hline

\multirow{2}{*}{01/14-06/16} & 0.0156& 0.0042& 0.0151&0.0038
\\
                           &(0.002 **) & (0.064 .) &
                           (0.0012 **) &(0.046 *)\\
\multirow{2}{*}{07/14-12/16} & -0.0039 &-0.0048&-0.0022&-0.003                                                                           \\
                            &(0.414)                                                      & (0.076 .)                                                         & (0.585)  &(0.096 .)                                                        \\
\multirow{2}{*}{01/15-06/17} & -0.2753&-0.109&-0.2144&-0.0481                                                                         \\
                           &(1.30E-06 ***)                                                      & (0.00016 ***)                                                         & (6.46E-06 ***) &(0.011 *)                                                         \\
\multirow{2}{*}{07/15-12/17} & -0.5422                                                      & -0.1448                                                           & -0.4691 &  -0.0717                                                      \\
                           &(2.80E-09 ***)                                                      & (0.0004 ***)                                                         & (1.56E-09 ***) &(0.009 **)                                                         \\
\multirow{2}{*}{01/16-06/18} & -0.7848 &-0.1155&-0.7355&-0.0661                                                                   \\
                           &(2.05E-013 ***)                                                      & (0.042 *)                                                         & (9.1E-16 ***)  &(0.061.)                                                        \\
\multirow{2}{*}{07/16-12/18} & -0.681 &0.0717&-0.7508&0.0018                                                                   \\
                           &(1.9E-12 ***)                                                      & (0.1303)                                                         & (4.49E-15 ***)  &(0.953)                                                        \\
                           
                           \bottomrule
\end{tabular}

Note: the p-value of the estimation of monthly TNC trips is shown in the bracket(. : $\leq$ 0.1; *: P $\leq$ 0.05; **: P $\leq$ 0.01; ***: P $\leq$ 0.001).
\end{table}

\highlighttext{Figure~\ref{overall} indicates the long-term examination might not be appropriate to understand the labor supply and revenue in taxi market due to their non-linear trend from 2014 to 2018. Thus, we testify the temporal variation based on the shorter time periods via OLS}. The results are presented in Table~\ref{tab:yellow taxi market time period variation} and Table~\ref{tab:green taxi market time period variation}. For yellow taxi drivers, the losses in total market revenue and labor supply (at 0.001 significance level) are found to be more significant than the losses at the individual level (at 0.05 significance level). The trend of the changes in monthly work hours of all taxi drivers has been observed to best echo the increasing trend in the number of TNC trips (as early as July 2013), followed by the market revenue. However, the drivers' monthly work hours are barely affected over time. The performances of the green taxi market at the beginning stage (January 2014) are different from that of the yellow taxi market. We observe more positive impacts in the green taxi market as the market revenue and total labor supply increase with positive coefficients, primarily due to the induced ridership outside Manhattan (e.g., there is 0.02\% increase in total labor supply of the green taxi market along with 1\% increase in the number of TNC trips). 

\added{The statistical estimates for both long-term and short-term taxi market performance directly reject the first hypothesis that the rise of TNC trips does not significantly impact the labor supply and revenue of the taxi market}. Our findings of the monthly income and monthly work hours for both yellow and green taxi markets lead to a counter-intuitive observation: individuals' work hours are barely affected by the significant decrease in their income. That observation contradicts the NS assumption, where taxi drivers are assumed to be revenue maximizers. One possible explanation for this observation is that the increased average minutes per taxi trip (positive coefficient at 0.001 and 0.05 significance level for yellow and green taxis) makes up for the reduced number of trips. Yet another plausible explanation is that the taxi drivers may have a specific reference target, and their work hours are no longer strictly positively related to their daily revenue. Following this explanation, taxi drivers may choose to decrease their income target and reach this target with lower wage rates and the same work hours. %This behavior can be illustrated in Figure~\ref{refer}, where the income target changes from $T^1$ to $T^2$, and the works hour keeps the same $H^*$, contrary to the decrease in work hours under NS as shown in Figure~\ref{neo}).
If this is the case, such an observation is indicative of the existence of RDP behavior in the current taxi market. And we next test the validity of such an explanation through statistical analyses. 

% \begin{figure}[!htb]
%      \centering
%      \includegraphics[width=0.5\linewidth]{fig/neo.png}
%      \caption{The work hours change based on the NS behavior}\label{neo}
% \end{figure}

% \begin{figure}[!htb]
%      \centering
%      \includegraphics[width=0.5\linewidth]{fig/refer.png}
%      \caption{The work hours change based on the RDP behavior}\label{refer}
% \end{figure}

\subsection{Taxi drivers' labor supply behavior and wage elasticity}
\added{While it is observed that the taxi market has been significantly affected by the rise of TNCs, we notice that taxi drivers' work hours do not change even though their income decreases. As taxi drivers are aware of the competition from the TNC sector, under the neo-classical scheme, it is reasonable to expect that taxi drivers will lower their work hours if their expected wage decreases, which obviously contradicts the aforementioned observation that their work hours are barely affected. With the increasing number of TNC trips, the underlying logic is that the taxi market share is expected to decrease and taxi drivers are likely to be more difficult to reach the same income level as before with the same amount of daily efforts (work hours) due to the competition. \highlighttext{In light of this counter-intuitive observation, we propose %the following research questions to investigate the reasons: \emph{\textbf{do the drivers decrease their expected wage along with the increasing number of TNC trips?} And 
the second hypothesis that \emph{\textbf{the increase of TNC trips does not decrease the taxi drivers' expected wage}}}. %Taxi drivers are likely to be more difficult to reach an income level as before with the same amount of daily efforts (work hours) due to the increase number of TNC trips.}
As discussed earlier, targeting behavior discourages the employee's motivation, lead to lower productivity, eventually conduct a vicious circle for the industry. This idea gives rise to our suspicion that the taxi labor supply may now be better explained by the RDP behavior instead of the neoclassical one. On this basis, our interest is to shed light on quantifying the RDP behavior in the taxi market to interpret the taxi drivers' non-intuitive response to the TNCs competition. This question has its roots in \highlighttext{Crawford and Meng's~\cite{crawford2011new}} study, where the taxi labor supply behavior is mainly driven by the income target. %To this end, the labor supply estimation under both the RDP and NS models can be analyzed from two aspects: wage proportion and wage elasticity. 
In this regard, we propose %the third research question: \emph{\textbf{is the RDP behavior present among taxi drivers with the increasing number of TNC trips? }} 
the third hypothesis that \emph{\textbf{taxi drivers do not present RDP behavior,}}}  and use the wage decomposition method to testify the labor supply behavior.

\highlighttext{We examine the second hypothesis using PLS to regress the drivers' average daily expected wage on the set of temporal indicators, the improvement surcharge indicators, and the log-transformation of monthly TNC trips.} The results are given in Table~\ref{tab:expected wage1} and Table~\ref{tab:expected wage2}. 

% \usepackage{multirow}
\begin{table}[!h]
\centering
\caption{Result of log-transformation of TNC trips: experiment \uppercase\expandafter{\romannumeral1} (PLS estimation)}
\label{tab:expected wage1}
\begin{tabular}{lcccc}
\toprule
\multirow{2}{*}{Month/Year}                                & \multicolumn{2}{c}{Yellow taxi} & \multicolumn{2}{c}{Green taxi} \\
\cmidrule(l){2-3} \cmidrule(l){4-5}
                                                                    & Coefficient     & P value  & Coefficient     & P value  \\
                                                                    \hline
\multirow{1}{*}{01/13-06/15} & 0.0009 & 0.457 &-&- \\
\multirow{1}{*}{07/13-12/15} & 0.0006& 0.622&-&-\\
\multirow{1}{*}{01/14-06/16} & -0.0005 & 0.698&0.0044&0.046 *\\
\multirow{1}{*}{07/14-12/16} & -0.0031& 0.085 .&-0.0050&0.044 *\\
\multirow{1}{*}{01/15-06/17} & -0.0518& 0.006 **&-0.1167&4.16E-06 ***\\
\multirow{1}{*}{07/15-12/17} & -0.0816& 0.001 ***&-0.1537 & 2.85E-05 ***\\
\multirow{1}{*}{01/16-06/18} & -0.0656& 0.069.&-0.1113 &0.028 *\\
\multirow{1}{*}{07/16-12/18} & 0.0665& 0.068.&0.0556&0.188\\
                           \bottomrule
\end{tabular}

Note: *: P $\leq$ 0.05; **: P $\leq$ 0.01; ***: P $\leq$ 0.001.

\end{table}

\begin{table}[!h]
\centering
\caption{Result of log-transformation of monthly TNC trips: experiment \uppercase\expandafter{\romannumeral2} (PLS estimation)}
\label{tab:expected wage2}
\begin{tabular}{lcccc}
\toprule
\multirow{2}{*}{Month/Year}                                & \multicolumn{2}{c}{Yellow taxi} & \multicolumn{2}{c}{Green taxi} \\
\cmidrule(l){2-3} \cmidrule(l){4-5}
                                                                    & Coefficient    & P value  & Coefficient    & P value  \\
                                                                    \hline
\multirow{1}{*}{01/13-12/14} &0.0013 &0.469 & -&-\\
\multirow{1}{*}{07/13-06/15} &0.0014 &0.293&-&-\\
\multirow{1}{*}{01/14-12/15} &0.0005 &0.761& 0.0045&0.073 .\\
\multirow{1}{*}{07/14-06/16} &-0.0020 &0.141&-0.0020&0.211 \\
\multirow{1}{*}{01/15-12/16} &-0.0650 &0.0002 ***&-0.0779&0.004 ** \\
\multirow{1}{*}{07/15-06/17} &-0.0401&0.0022 ** &-0.1447 & 0.003 **\\
\multirow{1}{*}{01/16-12/17} &-0.1218 & 0.005 **&-0.2065 &0.001 ***\\
\multirow{1}{*}{07/16-06/18} & 0.0323&0.512&0.0597&0.293\\
\multirow{1}{*}{01/17-12/18} & 0.1279&0.027 *&0.1900&0.003 **\\
                           \bottomrule
\end{tabular}

Note: *: P $\leq$ 0.05; **: P $\leq$ 0.01; ***: P $\leq$ 0.001.
\end{table}

\begin{figure}[!h]
\centering
\includegraphics[width=160mm]{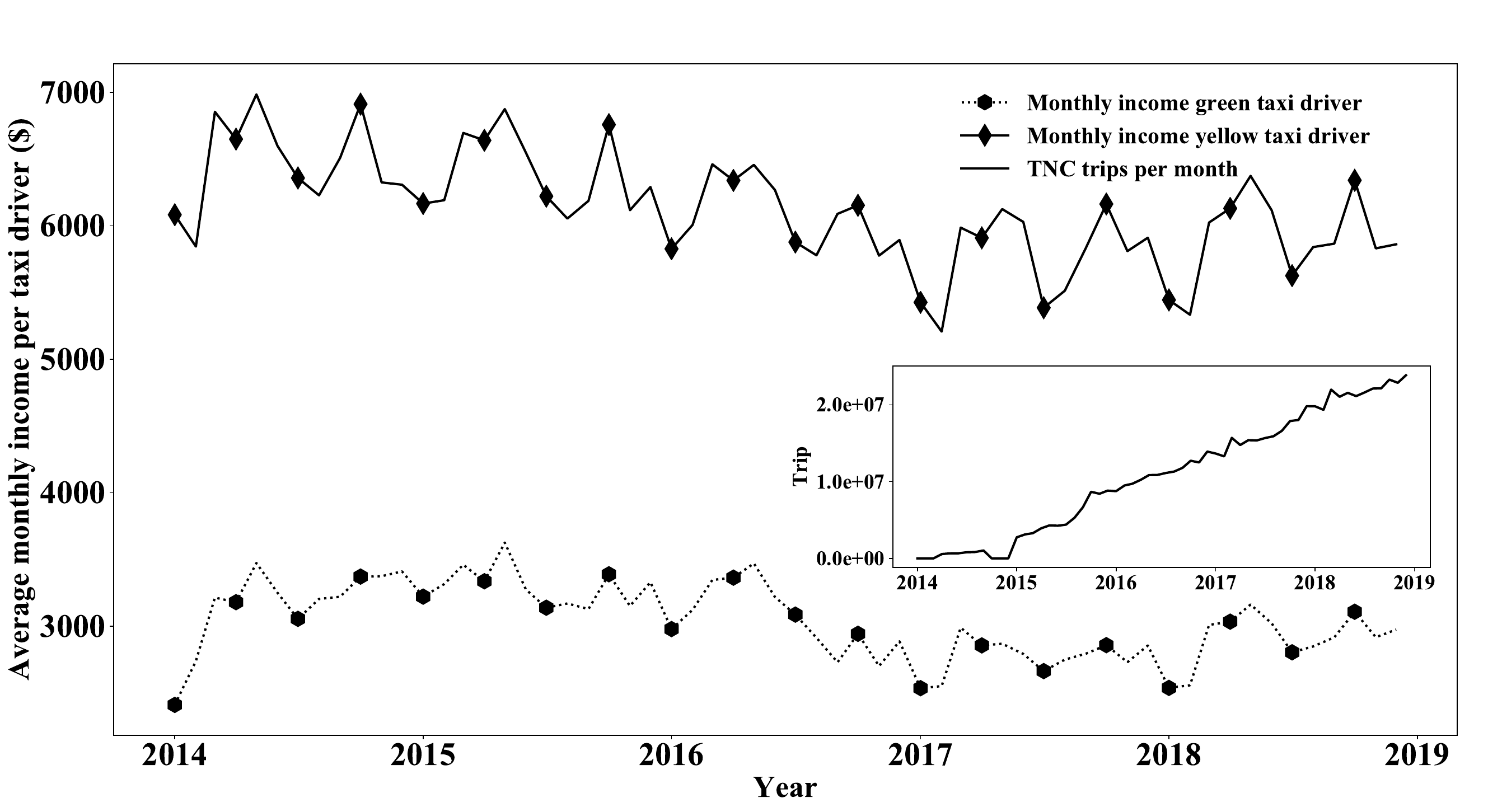}
\caption{Monthly TNC trips vs Monthly income per taxi driver}
\label{fig:monthly fare  & TNC trips}
\end{figure}

\begin{figure}[H]
\centering
\includegraphics[width=160mm]{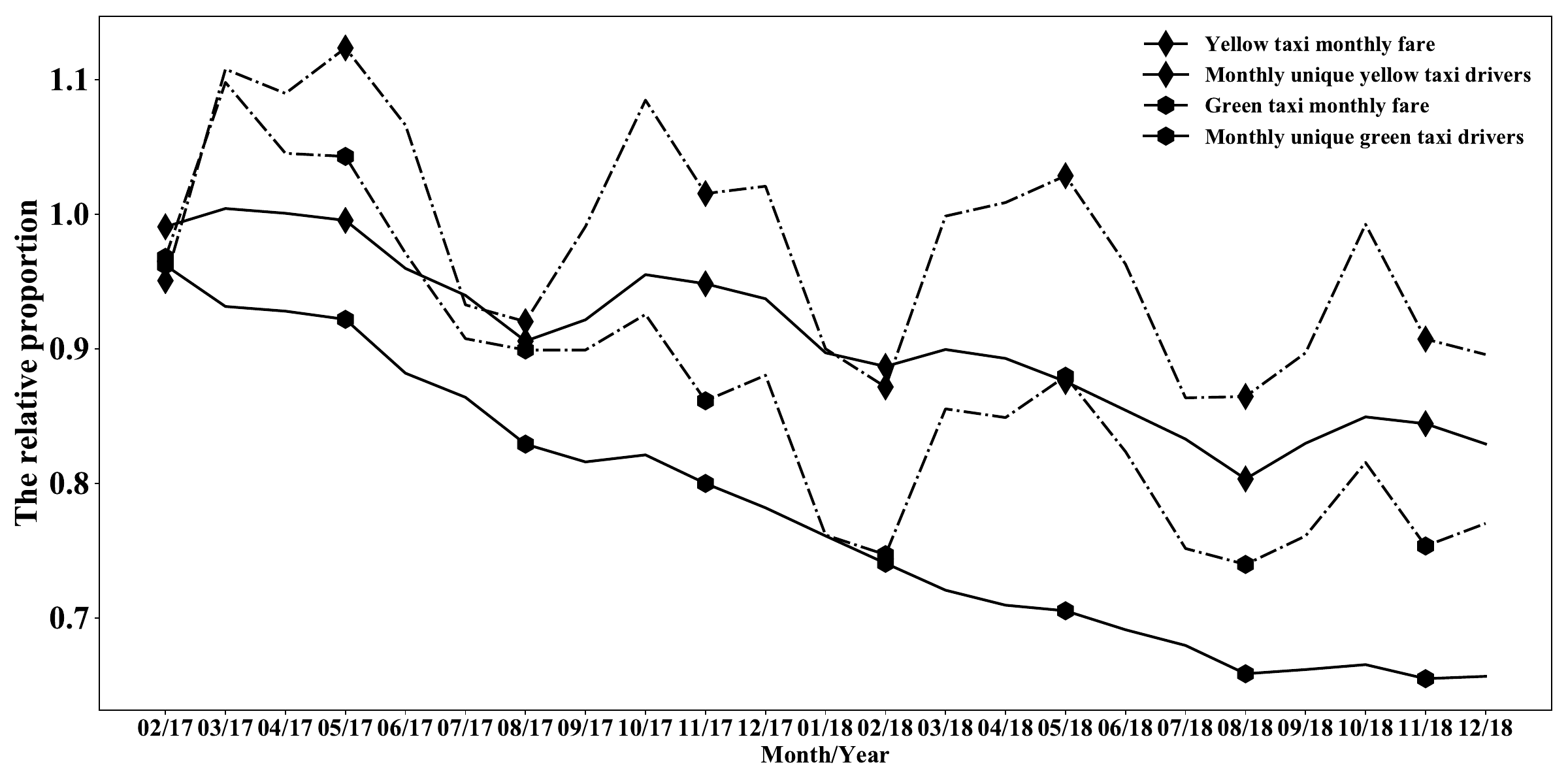}
\caption{The relative variation of fare and taxi drivers for yellow and green taxis} (base year: January 2017)
\label{fig:proportion}
\end{figure}

Although yellow taxi drivers' expected wage is non-significant related with month TNC trips from January 2013 to December 2014. Green taxi drivers are found to increase their expected wage due to the increase of TNC trips, which implies that the taxi market is still under-supplied with TNC trips in the beginning stage. Besides, the TNC trips are also found to negatively impact the expected wage from January 2015 to December 2017 for both green and yellow taxi drivers, which indicates the taxi market gradually shifted into the over-supplied state with increasing competition between taxis and TNCs (at 0.01 significance level). However, we observe that the increasing of TNC trips is found to be statistic significant and positively related to drivers' expected wage in 2018 and an individual-level income rebound presents during this period, as shown in Figure~\ref{fig:monthly fare  & TNC trips}. The income rebound is primarily due to the fact that the loss of total market supply (drivers quit the market) is faster than the reduction in taxi ridership and total market revenue, as shown in Figure ~\ref{fig:proportion}. As a consequence, the results from the experiment reject the second hypothesis that the increase of TNC trips does not decrease the taxi drivers' expected wage. Besides, Table~\ref{tab:expected wage1} and Table~\ref{tab:expected wage2} also indicate the consistency of our results under different data compositions. 

\highlighttext{To testify taxi driver's labor supply behavior in the third hypothesis, we measure the unanticipated and anticipated transitory wage variation based on the wage decomposition method}. Besides, we conduct the t-test to examine the change of the unanticipated transitory wage proportion in the current year group compared with its base year group. 
\begin{table}[!h]
\centering
\caption{Results of wage variation decomposition: experiment \uppercase\expandafter{\romannumeral1} (PLS estimation)}
\label{tab:wage decomposition1}
\begin{tabular}{lcccccc}
\toprule
\multirow{2}{*}{Month/Year}                                & \multicolumn{3}{c}{Yellow taxi} & \multicolumn{3}{c}{Green taxi} \\
\cmidrule(l){2-4} \cmidrule(l){5-7}
                                                                    & Fixed     & Anticipated  & Unanticipated   & Fixed     & Anticipated  & Unanticipated   \\
                                                                    \hline
\multirow{2}{*}{01/13-06/15}& 9.04E-06& 0.0021& 0.0002 &-&-&-\\
                           & (0.42\%)&(90.93\%)&(8.65\%)&-&-&-
\\

\multirow{2}{*}{07/13-12/15} & 1.46E-05& 0.0020& 0.0002&-&-&-
\\
                           & (0.65\%) & (90.05\%) &
                           (9.30\%) &-&-&-\\  
                           
\multirow{2}{*}{01/14-06/16}& 0.0001& 0.00198& 0.0002  &0.000412&0.0049&0.0004 \\
                           & (4.42\%)&(85.64\%)&(9.94\%)&(7.26\%)&(86.42\%)&(6.32\%)
\\
\multirow{2}{*}{07/14-12/16} & 0.0002& 0.0018& 0.0003& 0.0002&0.0040&0.0006 
\\
                           & (9.42\%) & (78.45\%) &
                           (12.13\%)& (3.22\%) & (84.36\%) &
                           (12.42\%)  \\  
\multirow{2}{*}{01/15-06/17} & 0.0002                                                       & 0.0029                                                          & 0.0004 &     3.16E-30&0.0070&0.0011                                                    \\
                           & (5.62\%)                                                      & (82.09\%)                                                         & (12.29\%)  & (0\%)                                                      & (85.98\%)                                                         & (14.02\%)                                                        \\
\multirow{2}{*}{07/15-12/17 } & 6.04E-05                                                       & 0.0025                                                          & 0.0005&7.89E-31& 0.0062&0.0091                                                    \\
                           & (1.95\%)                                                      & (80.65\%)                                                         & (17.40\%)  & (0\%)                                                      & (87.24\%)                                                         & (12.76\%)                                                         \\

\multirow{2}{*}{01/16-06/18} & 0.0002                                                      & 0.0025                                                           & 0.0005  &7.10E-30& 0.0055&0.0013                                                \\
                           & (6.76\%)                                                      & (77.28\%)                                                         & (15.96\%)  & (0\%)                                                      & (80.86\%)                                                         & (19.14\%)                                                         \\
\multirow{2}{*}{07/16-12/18} & 5.82E-05                                                       & 0.0021                                                          & 0.0004 &    3.16E-30&0.0028&0.0008                                                    \\
                           & (2.33\%)                                                      & (83.33\%)                                                         & (14.34\%)  & (0\%)                                                      & (78.03\%)                                                         & (21.97\%)                                                         \\
                           \bottomrule
\end{tabular}

Note: the proportion of variation of each part in the total variation is in the bracket.
\end{table}

\begin{table}[!h]
\centering
\caption{Results of wage variation decomposition: experiment \uppercase\expandafter{\romannumeral2} (PLS estimation)}
\label{tab:wage decomposition2}
\begin{tabular}{lcccccc}
\toprule
\multirow{2}{*}{Month/Year}                                & \multicolumn{3}{c}{Yellow taxi} & \multicolumn{3}{c}{Green taxi} \\
\cmidrule(l){2-4} \cmidrule(l){5-7}
                                                                    & Fixed     & Anticipated  & Unanticipated   & Fixed     & Anticipated  & Unanticipated   \\
                                                                    \hline

\multirow{2}{*}{01/13-12/14 }& 9.6E-07& 0.0022& 0.0002 &-&-&-\\
                           & (0.04\%)&(92.84\%)&(7.12\%)&-&-&-
\\
\multirow{2}{*}{07/13-06/15} & 4.03E-05& 0.0021& 0.0002 &-&-&- \\
                           & (1.76\%)&(91.50\%)&(6.74\%)&-&-&-
\\
\multirow{2}{*}{01/14-12/15} & 4.65E-05& 0.0020& 0.0002&0.0005&0.0056&0.0003
\\
                           & (2.05\%) & (89.80\%) &
                           (8.15\%)& (8.29\%) & (87.17\%) &
                           (4.54\%)\\

\multirow{2}{*}{07/14-06/16} & 5.08E-05                                                       & 0.0015                                                           & 0.0002 &   7.15E-07&0.0018&0.0002                                                        \\
                           & (3.16\%)                                                      & (85.10\%)                                                         & (11.98\%) & (0.04\%)                                                      & (91.54\%)                                                         & (8.42\%)                                                           \\
\multirow{2}{*}{01/15-12/16} & 9.34E-05                                                       & 0.0020                                                           & 0.0003  & 3.16E-30&0.0047&0.0006                                                          \\
                           & (3.93\%)                                                      & (82.97\%)                                                         & (13.10\%) & (0\%)                                                      & (88.06\%)                                                         & (11.94\%)                                                          \\
\multirow{2}{*}{07/15-06/17} & 1.21E-05                                                       & 0.0024                                                          & 0.0005    & 7.89E-31&0.0066&0.0007                                                   \\
                           & (0.41\%)                                                      & (83.33\%)                                                         & (16.26\%)    & (0\%)                                                      & (89.84\%)                                                         & (10.16\%)                                                        \\
\multirow{2}{*}{01/16-12/17} & 8.97E-05                                                      & 0.0022                                                          & 0.0006    & 7.10E-30&0.0054&0.0012                                                       \\
                           & (3.10\%)                                                      & (76.90\%)                                                         & (20\%)& (0\%)                                                      & (82.05\%)                                                         & (17.95\%)                                                           \\
\multirow{2}{*}{07/16-06/18} & 7.05E-05                                                       & 0.0022&0.0005   &3.16E-05&0.0031&0.0008                                                                                                      \\
                           & (2.52\%)                                                      & (79.30\%)                                                         & (18.18\%)  & (0.80\%)                                                      & (78.68\%)                                                         & (20.52\%)                                                         \\
\multirow{2}{*}{01/17-12/18} & 0.0001                                                      & 0.0024                                                         & 0.0004    & 3.16E-30&0.0030&0.0009                                                 \\
                           & (4.35\%)                                                      & (81.33\%)                                                         & (14.32\%) & (0\%)                                                      & (76.55\%)                                                         & (23.45\%)                                                          \\
                                                  
                           \bottomrule
\end{tabular}

Note: the proportion of variation of each part in the total variation is in the bracket.
\end{table}

\begin{figure}[!h]
  \centering
  \includegraphics[width=170mm]{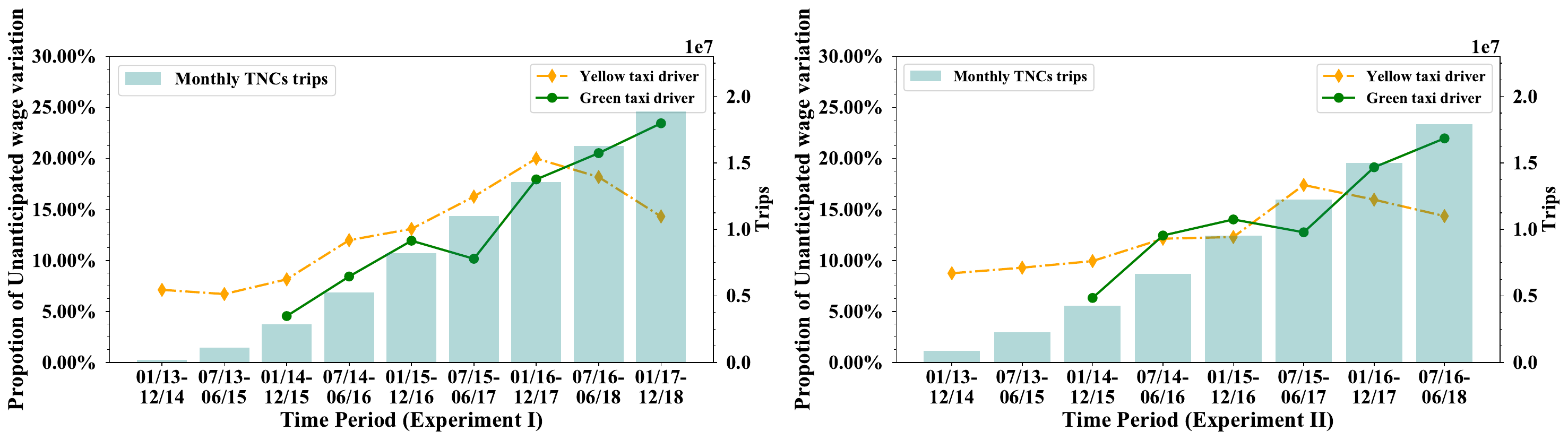}
\caption{Relationship between unanticipated wage variation and monthly TNC trips}
\label{fig:relationship}
\end{figure}

The wage decomposition results are presented in Table~\ref{tab:wage decomposition1} and Table~\ref{tab:wage decomposition2}. The results quantify the fixed, anticipated, and unanticipated transitory wage variation of drivers' wage rate over time. For the yellow taxi drivers, we observe that the anticipated transitory variation dominates the total variation in both experiments. The proportion of unanticipated wage variation (8\%) from January 2013 to December 2014 is close to the findings in Farber's study~\cite{farber2015you}, where the unanticipated hourly transitory wage variation is reported to be 12.1\% when the taxi market is still monopolistic. \highlighttext{That is, when supply is not saturated, almost all of the taxi drivers' work behavior can be explained by the NS behavior.} However, the proportion of anticipated wage variation is observed to decrease as TNCs grow. The unanticipated transitory wage variation reaches its peak at the end of 2017 when 20 \% of yellow taxi drivers' behavior can be explained by RDP as shown in Figure~\ref{fig:relationship}. This result matches well with our previous findings when drivers decrease their expected wage as shown in Table~\ref{tab:expected wage1} and Table~\ref{tab:expected wage2}, as well as when the market revenue and labor supply significantly decreased (see Table~\ref{tab:yellow taxi market time period variation}). For green taxi drivers, only a small proportion (6\%) of green taxi drivers shows RDP behavior in the beginning stage (from January 2014 to June 2016). With the increasing competition from TNCs (see Figure~\ref{overall}), green taxi drivers' expected wage decreases, which results in the reduction of green taxi drivers in their work hours and income from July 2015 to June 2017 (see Table~\ref{tab:green taxi market time period variation}). Therefore, green taxi drivers exhibit revenue optimizing behavior as suggested by the NS (drivers having RDP behavior also slightly decrease during this period from 14\% to 12\% in experiment \uppercase\expandafter{\romannumeral1} and 12\% to 10\% in experiment \uppercase\expandafter{\romannumeral2}). After the second half of 2017, green taxi drivers show increasing RDP behavior due to the increasing competition from TNCs (see Figure~\ref{fig:relationship}). Finally, we observe that RDP behavior captures over 20\% of the green taxi drivers in the current taxi market.

\begin{figure}[!h]
\centering
\includegraphics[width=160mm]{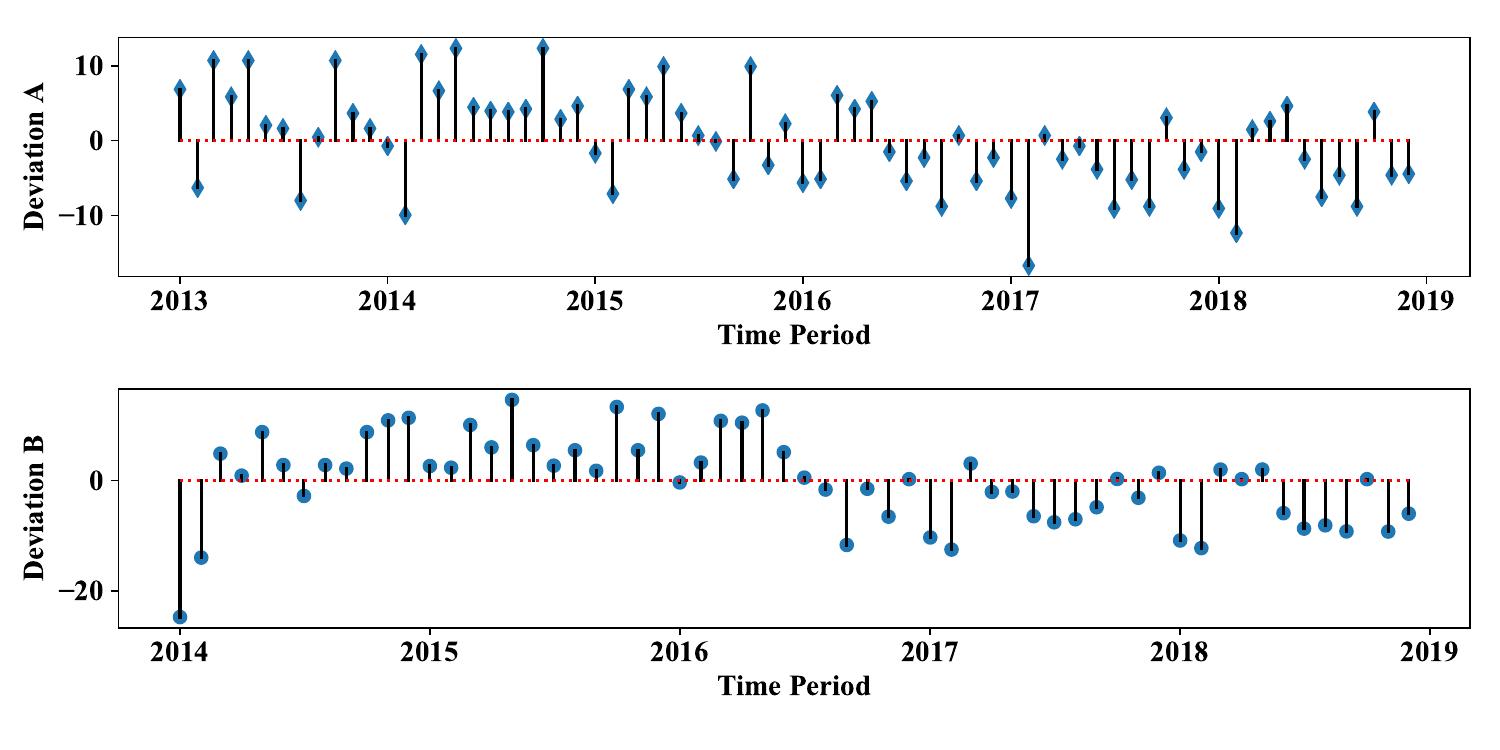}
\caption{Monthly work hours per driver (A: yellow taxi mean =171.5 h; B: green taxi mean=129.7 hours)}
\label{fig:hours}
\end{figure}

The t-test in Table~\ref{tab:test hypothesis 2 for exp 1} and Table ~\ref{tab:test hypothesis 2 for exp 2} show a significant higher unanticipated transitory wage for both yellow and green taxis than their base year. The relationship between TNC trips and unanticipated transitory wage variation can be seen in Figure~\ref{fig:relationship}. The results indicate that drivers are facing more uncertainty in earning opportunities and clearly illustrate the change of drivers' labor supply behavior over time due to the increasing number of TNC trips. Therefore, \added{we reject the third hypothesis and conduct that RDP behavior presents among taxi drivers with the increasing number of TNC trips}. Based on the results, we observe that yellow taxi drivers face much more serious competition than green taxi drivers before 2017. Moreover, such competition leads to an unsustainable state in the ride-sharing market and results in 20\% yellow taxi drivers having RDP behavior at the end of 2017, which means a high proportion of taxi drivers lose confidence in the taxi industry. Finally, yellow taxi drivers quit the taxi market. Besides, the green taxi market benefits from the increase in demand at the beginning of TNCs' growth. Meanwhile, green taxi drivers present NS behavior before July 2017. However, the unanticipated transitory wage variation of green taxi drivers is found to increase after June 2017 and account for over 20\% of the total wage variation in both experiments at the end of 2018, which is over three times as compared to when the taxi market is still monopolistic. Consequently, the RDP behavior should not be ignored, and NS behavior is no longer suitable to interpret the total taxi drivers' work behavior in a competitive market. Instead, at least $20\%$ of green taxi drivers perform in a loss-aversion manner in the market rather than the revenue-maximizing behavior, which is widely used when the taxi market is still monopolistic. This finding is aligned with the second explanation for the question that we raised to the OLS model. That is, the driver has a specific reference target. Moreover, the fact that individual labor supply is found to be barely unaffected (see Figure~\ref{fig:hours}) while their monthly income is found to be significantly decreased can be explained by the co-existence of NS and the RDP behavior in the market. Furthermore, combining the examinations of the second and third hypotheses, we conclude that drivers decrease their income target and some of them even quit the market, so that the remaining drivers are observed to still serve the same amount of work hours. It points out the necessity to consider the regulation of TNCs for the sustainability of the taxi market. This issue requires further investigation with related data sources. Finally, Table~\ref{tab:wage decomposition1} and Table~\ref{tab:wage decomposition2} again confirm the consistency of our results under different data compositions.

\begin{table}[!h]
\centering
\caption{t-test for the change of unanticipated wage variation in experiment\uppercase\expandafter{\romannumeral 1} }
\label{tab:test hypothesis 2 for exp 1}
\begin{tabular}{lcc}
\toprule
Month/Year   & Yellow taxi& Green taxi \\  
                                                                    \hline

01/13-06/15&0.896&-
\\
07/13-12/15&0.853&-
\\
01/14-06/16&0.696&0.957
\\

07/14-12/16  &         0.228&0.032 *                                            \\
01/15-06/17           &0.118     &0.0007 ***   \\
07/15-12/17  &0.038*&0.0047 ** \\
01/16-06/18 & 0.023 * &0.002 **\\
07/16-12/18   &0.051 .&0.003 *  \\
\bottomrule

\end{tabular}
\end{table}

\begin{table}[!h]
\centering
\caption{t-test for the change of unanticipated wage variation in experiment \uppercase\expandafter{\romannumeral 2} }
\label{tab:test hypothesis 2 for exp 2}
\begin{tabular}{lcc}
\toprule
Month/Year   & Yellow taxi& Green taxi \\  
                                                                    \hline

07/13-06/15&0.6461&-
\\
01/14-12/15&0.922&-
\\

07/14-06/16 &         0.367&0.606                                         \\
01/15-12/16           &0.121     &0.019 *   \\
07/15-06/17   &0.0104*&0.015 *  \\
01/16-12/17 & 0.048 * &0.002 **\\
07/16-06/18   &0.138&0.046 *\\
01/17-12/18& 0.195 &0.006 **\\
\bottomrule

\end{tabular}

\end{table}

\begin{table}[!h]
\centering
\caption{Wage elasticity of taxi labor supply: experiment \uppercase\expandafter{\romannumeral 1} (PLS estimation)}
\label{tab:elast1}
\begin{tabular}{lcccc}
\toprule
\multirow{2}{*}{Month/Year}& \multicolumn{2}{c}{Yellow taxi}&\multicolumn{2}{c}{Green taxi}\\
\cmidrule(l){2-3} \cmidrule(l){4-5}
     & Coefficient & $Adj.R^2$ & Coefficient & $Adj.R^2$ \\
\hline
\multirow{2}{*}{01/13-06/15} & 0.6611      & 0.844& -&-\\
          & (5.00E-13 ***)     &   &-&        \\
\multirow{2}{*}{07/13-12/15} & 0.6526      & 0.844&-&-\\
          & (5.06E-13 ***)     &         &-&     \\
\multirow{2}{*}{01/14-06/16} & 0.6383      & 0.837&0.8006&0.92\\
          & (9.39E-13 ***)     &    &(4.55E-07 ***)&          \\
\multirow{2}{*}{07/14-12/16} & 0.5592      & 0.731&0.6228&0.953\\
          & (1.10E-09 ***)     &        &(5.96E-08 ***)&       \\
\multirow{2}{*}{01/15-06/17} & 0.5040      & 0.764&0.5822&0.903\\
          & (1.69E-10 ***)     &     &(7.51E-13 ***)&         \\
\multirow{2}{*}{07/15-12/17} & 0.4780      & 0.712&0.6103&0.909\\
          & (2.82E-09 ***)     &  &(2.48E-16 ***)&             \\
\multirow{2}{*}{01/16-06/18} & 0.4924      & 0.745&0.5770&0.863\\
          & (5.23E-10 ***)     &      &(8.34E-14 ***)&        \\
\multirow{2}{*}{07/16-12/18}& 0.4944      & 0.699&0.4776&0.541\\
          & (5.49E-09 ***)     &     &(2.24E-06 ***)&          \\
          \bottomrule
          
\end{tabular}

Note: the p-value of the estimate for log-transformation of monthly income per taxi driver is in the bracket (. : p $\leq$ 0.1; *: P $\leq$ 0.05; **: P $\leq$ 0.01; ***: P $\leq$ 0.001).
\end{table}

\begin{table}[!h]
\centering
\caption{Wage elasticity of taxi labor supply: experiment \uppercase\expandafter{\romannumeral 2} (PLS estimation)}
\label{tab:elast2}
\begin{tabular}{lcccc}
\toprule
\multirow{2}{*}{Month/Year}& \multicolumn{2}{c}{Yellow taxi}&\multicolumn{2}{c}{Green taxi}\\
\cmidrule(l){2-3} \cmidrule(l){4-5}
     & Coefficient & $Adj.R^2$ & Coefficient & 
     $Adj.R^2$ \\
\hline
\multirow{2}{*}{01/13-12/14}
& 0.6477      & 0.762&-&-\\
          & (5.68E-11 ***)     &  &-&          \\
\multirow{2}{*}{07/13-06/15} & 0.6582      & 0.854&-&-\\
          & (7.09E-11 ***)     &&-&            \\
\multirow{2}{*}{01/14-12/15} & 0.6436      & 0.827&0.8095&0.929\\
          & (4.76E-10 ***)     & &(2.35E-14 ***) &          \\
\multirow{2}{*}{07/14-06/16} & 0.6243      & 0.772&0.6834&0.762\\
          & (1.02E-08 ***)     & &(1.59E-08 ***) &            \\
\multirow{2}{*}{01/15-12/16}  & 0.5362      & 0.702&0.6087&0.862\\
          & (1.99E-07 ***)     & &(3.74E-11 ***)  &           \\
\multirow{2}{*}{07/15-06/17} & 0.5151      & 0.74&0.6379&0.933\\
          & (4.24E-08 ***)     &   &(1.30E-14 ***) &           \\
\multirow{2}{*}{01/16-12/17} & 0.4746      & 0.687&0.5960&0.893\\
          & (3.38E-07 ***)     &  &(2.33E-12 ***)            \\
\multirow{2}{*}{07/16-06/18} & 0.488      & 0.696 &0.5376&0.719\\
          & (2.45E-07 ***)     &&(1.02E-07 ***) &              \\
\multirow{2}{*}{01/17-12/18} &0.5243      & 0.798&0.4523&0.515\\
          & (2.59E-09 ***)     & &(4.81E-05 ***) &             \\
          \bottomrule
          
\end{tabular}

Note: the p-value of the estimate for log-transformation of monthly income per taxi driver is in the bracket (. : p $\leq$ 0.1; *: P $\leq$ 0.05; **: P $\leq$ 0.01; ***: P $\leq$ 0.001).

\end{table}

\begin{figure}[!h]
\centering
\includegraphics[width=170mm]{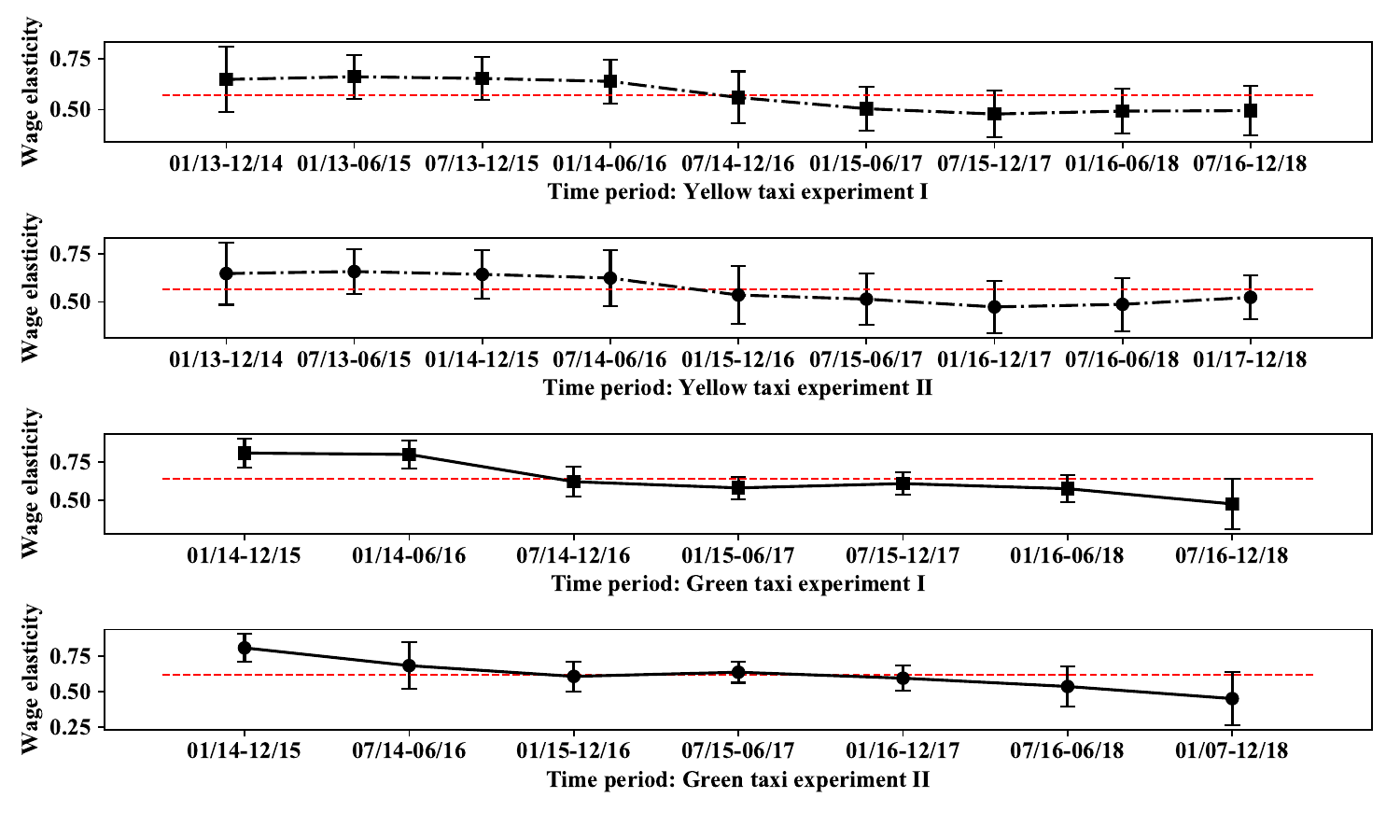}
\caption{Estimated wage elasticity}
\label{fig:elasticity}
\end{figure}

\highlighttext{Finally, although the increase in unanticipated transitory wage variation suggests that drivers' behavior may shift from NS to RDP, it is not equivalent to assert that the RDP should explain drivers' behavior. We further conduct the wage elasticity analysis to provide a better understanding of this issue. The results are presented in Table~\ref{tab:elast1} and Table~\ref{tab:elast2}. We observe that wage elasticity yields a similar trend as the change of unanticipated transitory wage variation proportion and the change of RDP behavior for both yellow and green taxi drivers (see Figure~\ref{fig:relationship}). The wage elasticity for both yellow and green taxi drivers remains positive, which is inconsistent with the elasticity of RDP models being -1. Although the wage elasticity is positive, it implies the yellow taxi drivers reached the lowest wage elasticity at the period of January of 2016 to December 2017 (see Figure~\ref{fig:elasticity}), when 1\% increase in drivers' wage rate leads to 0.47\% increase in monthly work hours. Besides, the wage elasticity of yellow taxi stayed above 0.6 before June 2016 but dropped rapidly until December 2017. This finding corresponds to when the transitory wage variation proportion has changed, as shown in Figure~\ref{fig:relationship}. For the wage elasticity of green taxi drivers, there is a rebound at the period of July 2015 to June 2017 (see Figure~\ref{fig:elasticity}), which confirms our insights that they show revenue-maximizing behavior along with the increasing of TNC trips during this period. Since then, the wage elasticity has decreased. The similar results from the comparison of the two experiments again verify the consistency of our results.}

\added{In conclusion, the results from both the wage variation decomposition and wage elasticity of labor supply are sufficient to reject the third hypothesis and confirm that taxi drivers show RDP behavior.} The labor supply behavior of both yellow and green taxi drivers has changed since the increase of TNC trips at the beginning of 2015. At the end of study period, RDP explains over 14\% yellow taxi drivers and 20\% green taxi drivers' behavior. The RDP behavior among taxi drivers implies that an increasing number of drivers would quit the taxi market due to the loss of confidence~\cite{eliaz2014reference}. And the slightly weakened RDP behavior among yellow taxis after 2018 is likely to support this claim, where the taxi market has lost a number of active taxi drivers so that the remaining drivers are less pessimistic about the market with less competition from the same sector. The insights from both labor supply estimation and wage elasticity analyses suggest that taxi drivers show increasingly negative responses to the market over time. \highlighttext{Since pricing regulation and market entry regulation serve as one fundamental tool in reshaping drivers' labor supply behavior and maintain the balance between taxis and FHVs. Taking labor supply behavior into consideration is crucial to promote a sustainable and equitable environment in the competitive market of taxis and FHVs.}

\section{Conclusion}

The taxi industry has undergone tremendous changes in recent years due to the rapid growth of TNCs. This results in losses of ridership, labor supply, and asset value in the taxi market. \added{To explore the impacts of TNCs on the taxi market performance metrics and taxi drivers' labor supply behavior, we propose three research questions and the corresponding hypotheses. The analyses for three research questions further confirm that the rise of TNCs not only decreases in the market labor supply and revenue but significantly changes the work behavior of taxi drivers}, and such change is likely to continue with increasing competition from TNCs. The major findings of our study can be summarized in three aspects.

First, we investigate the impact of TNCs on the overall revenue and labor supply of taxi market for the first research question. We find that while the rise of TNCs takes over the market share from the taxi industry in Manhattan, it also contributes to solving the under-supply issue in other boroughs of NYC. Besides, the investigation of the yellow taxi market performance shows that 1\% increase in TNC trips leads to 0.02\% loss in monthly market revenue and 0.01\% loss in the total market labor supply. Despite individual monthly work hours being barely affected, the increase in the number of TNC trips results in the reduction of total market labor supply and individual wage expectations.

Second, \added{to investigate the contradictory observations under the NS assumption, we propose the second and third research questions. The second research question addresses whether drivers' expected wage has been decreased due to the rise of TNC trips. The results suggest that even taxi drivers' expected wage remained unaffected before June 2016 for both green taxis and yellow taxis. Their expected wage decreased when the TNC trips grow rapidly. For this reason, taxi drivers can still reach the expectation-based income despite they pay the same work hours. Moreover, this insight sheds light on the existence of RDP behavior. The third research question is proposed to explore whether RDP behavior is present among taxi drivers and how much of the drivers present RDP behavior over the years. This question directly implies the lack of the taxi drivers' confidence in earning opportunities in the current market and the potential to quit the taxi market.} 

Third, we identify that the income target is the main factor that determines taxi drivers' labor supply behavior in the competitive market. In this domain, we apply the wage decomposition method to investigate the third research question and verified by the wage elasticity estimation. The significant increase in unanticipated wage variation shifts the drivers' labor supply behavior from NS to RDP with increasing competition from TNCs---this change occurred between July 2015 to June 2018 for yellow taxi drivers and after June 2016 for the green taxi drivers. The fact of 20\% RDP behavior among yellow taxi drivers at the end of 2017 indicates that the drivers quit the market due to the loss of confidence in the taxi market as RDP behavior increases. \highlighttext{\added{Over 14\% of yellow taxi drivers and 20\% of green taxi drivers presenting RDP behavior (four times than when the taxi market is still monopolistic) imply the low productivity and quantify the potential loss of taxi drivers. Furthermore, the estimated results from both wage variation and wage elasticity favor the co-existence of the NS and RDP behavior among taxi drivers, and supplement the existing literature for either supporting NS assumption or RDP assumption for drivers' labor supply behavior. Besides, it points out that the NS behavior is dominated at the TNCs' beginning stage. More importantly, it highlights the transition of individual behavior from NS to RDP due to the impact of TNCs and an unsustainable competition between taxis and TNCs. This imposes the need to reconsider entry regulations and pricing policies for the ride-hailing industry. }}

Several future directions are worth exploring based on the findings of the study. The first direction is to quantify the changes in the taxi drivers' target income due to the impact of TNCs, which can be conducted by using medallion transaction data along with the taxi and TNC trips data to gain insights on the net loss of taxi drivers with increasing competitions. In addition, it is meaningful to extend the research framework in our study to investigate the change in the labor supply of the taxi industry in other major cities around the world. While different cities are experiencing different levels of competition from TNCs, the joint analyses will lead to universal and unbiased understandings of the impact of TNC trips on the traditional taxi industry. Furthermore, the market equilibrium between demand and supply of rider sharing market may need to be revisited in light of the competition of multiple service providers under RDP behavior among the drivers.

\section*{Reference}
\bibliographystyle{unsrt}
%% ***   Set the bibliography file.   ***
\bibliography{main}

\begin{thebibliography}{10}

\bibitem{hall2018uber}
Jonathan~D Hall, Craig Palsson, and Joseph Price.
\newblock Is uber a substitute or complement for public transit?
\newblock {\em Journal of Urban Economics}, 108:36--50, 2018.

\bibitem{qian2020impact}
Xinwu Qian, Tian Lei, Jiawei Xue, Zengxiang Lei, and Satish~V Ukkusuri.
\newblock Impact of transportation network companies on urban congestion:
  Evidence from large-scale trajectory data.
\newblock {\em Sustainable Cities and Society}, page 102053, 2020.

\bibitem{lai2020evaluating}
Xiongfei Lai, Jing Teng, and Lu~Ling.
\newblock Evaluating public transportation service in a transit hub based on
  passengers energy cost.
\newblock In {\em 2020 IEEE 23rd International Conference on Intelligent
  Transportation Systems (ITSC)}, pages 1--7. IEEE, 2020.

\bibitem{lai2020resilient}
Xiongfei Lai, Jing Teng, Paul Schonfeld, and Lu~Ling.
\newblock Resilient schedule coordination for a bus transit corridor.
\newblock {\em Journal of Advanced Transportation}, 2020:1--12, 2020.

\bibitem{ukkusuri2020performance}
Satish Ukkusuri, Lu~Ling, Tho~V Le, and Wenbo Zhang.
\newblock Performance of right-turn lane designs at intersections.
\newblock 2020.

\bibitem{ling2023influencing}
Lu~Ling, Wenbo Zhang, Jie Bao, and Satish~V Ukkusuri.
\newblock Influencing factors for right turn lane crash frequency based on
  geographically and temporally weighted regression models.
\newblock {\em Journal of Safety Research}, 2023.

\bibitem{ling2017reliable}
Lu~Ling and Feng Li.
\newblock Reliable feeder bus schedule optimization in a multi-mode transit
  system.
\newblock In {\em 2017 IEEE 20th International Conference on Intelligent
  Transportation Systems (ITSC)}, pages 738--744. IEEE, 2017.

\bibitem{Dan2018}
Dan Rivoli and Erin Durkin.
\newblock {City Sets aside Sale of Taxi Medallions as Their Value Plummets}.
\newblock
  \url{http://www.nydailynews.com/new-york/city-sets-sale-taxi-medallions-plummets-article-1.3982966},
  2018.
\newblock [Online; accessed 10-July-2018].

\bibitem{timmurphy.org}
Tom Corrigan.
\newblock {San Francisco's Biggest Taxi Operator Seeks Bankruptcy Protection}.
\newblock
  \url{https://www.wsj.com/articles/san-franciscos-biggest-taxi-operator-seeks-bankruptcy-protection-1453677177/},
  2016.
\newblock [Online; accessed 24-Jan-2016].

\bibitem{Barry2019}
Barry Ritholtz.
\newblock Taxi medallion owners demand a bailout.
\newblock \url{https://ritholtz.com/2019/10/taxi-owner-bailouts/}, 2019.
\newblock [Online; accessed 17-October-2019].

\bibitem{WF2015}
Emma Whitford.
\newblock Greenpoint's growing taxi graveyard.
\newblock
  \url{http://gothamist.com/2015/08/21/why_yellow_cabs_are_taking_up_all_o.php#photo-1},
  2015.
\newblock [Online; accessed 21-Aug-2015].

\bibitem{Nikita2018}
Nikita Stewart and Luis Ferr\'{e}-Sadurn\'{i}.
\newblock Another taxi driver in debt takes his life. that's 5 in 5 months.
\newblock
  \url{https://www.nytimes.com/2018/05/27/nyregion/taxi-driver-suicide-nyc.html},
  2018.
\newblock [Online; accessed 27-May-2018].

\bibitem{contreras2017effects}
Seth~D. Contreras and Alexander Paz.
\newblock The effects of ride-hailing companies on the taxicab industry in las
  vegas, nevada.
\newblock {\em Transportation Research Part A: Policy and Practice}, 115:63 --
  70, 2018.

\bibitem{cetin2013economic}
Tamer Cetin and Kadir~Yasin Eryigit.
\newblock The economic effects of government regulation: Evidence from the new
  york taxicab market.
\newblock {\em Transport Policy}, 25:169--177, 2013.

\bibitem{ling2019forecasting}
Lu~Ling, Xiongfei Lai, and Li~Feng.
\newblock Forecasting the gap between demand and supply of e-hailing vehicle in
  large scale of network based on two-stage model.
\newblock In {\em 2019 IEEE Intelligent Transportation Systems Conference
  (ITSC)}, pages 3880--3885. IEEE, 2019.

\bibitem{qian2017time}
Xinwu Qian and Satish~V Ukkusuri.
\newblock Time-of-day pricing in taxi markets.
\newblock {\em IEEE Transactions on Intelligent Transportation Systems},
  18(6):1610--1622, 2017.

\bibitem{asamer2016optimizing}
Johannes Asamer, Martin Reinthaler, Mario Ruthmair, Markus Straub, and Jakob
  Puchinger.
\newblock Optimizing charging station locations for urban taxi providers.
\newblock {\em Transportation Research Part A: Policy and Practice},
  85:233--246, 2016.

\bibitem{frechette2019frictions}
Guillaume~R Frechette, Alessandro Lizzeri, and Tobias Salz.
\newblock Frictions in a competitive, regulated market: Evidence from taxis.
\newblock {\em American Economic Review}, 109(8):2954--92, 2019.

\bibitem{zha2016economic}
Liteng Zha, Yafeng Yin, and Hai Yang.
\newblock Economic analysis of ride-sourcing markets.
\newblock {\em Transportation Research Part C: Emerging Technologies},
  71:249--266, 2016.

\bibitem{qian2017taxi}
Xinwu Qian and Satish~V Ukkusuri.
\newblock Taxi market equilibrium with third-party hailing service.
\newblock {\em Transportation Research Part B: Methodological}, 100:43--63,
  2017.

\bibitem{farber2015you}
Henry~S. Farber.
\newblock {Why You Can't Find a Taxi in the Rain and Other Labor Supply Lessons
  from Cab Drivers}.
\newblock {\em The Quarterly Journal of Economics}, 130(4):1975--2026, 07 2015.

\bibitem{yang1998network}
Hai Yang and Sze~Chun Wong.
\newblock A network model of urban taxi services.
\newblock {\em Transportation Research Part B: Methodological}, 32(4):235--246,
  1998.

\bibitem{farber2008reference}
Henry~S Farber.
\newblock Reference-dependent preferences and labor supply: The case of new
  york city taxi drivers.
\newblock {\em American Economic Review}, 98(3):1069--82, 2008.

\bibitem{crawford2011new}
Vincent~P. Crawford and Juanjuan Meng.
\newblock New york city cab drivers' labor supply revisited:
  Reference-dependent preferences with rational-expectations targets for hours
  and income.
\newblock {\em American Economic Review}, 101(5):1912--32, August 2011.

\bibitem{eliaz2014reference}
Kfir Eliaz and Ran Spiegler.
\newblock Reference dependence and labor market fluctuations.
\newblock {\em NBER macroeconomics annual}, 28(1):159--200, 2014.

\bibitem{bewley2009wages}
Truman~F Bewley and Truman~F Bewley.
\newblock {\em Why wages don't fall during a recession}.
\newblock Harvard university press, 2009.

\bibitem{ling2018analyzing}
Lu~Ling, Feng Li, and Linhui Cao.
\newblock Analyzing the relationship between urban macroeconomic development
  and transport infrastructure system based on neural network.
\newblock In {\em Green Intelligent Transportation Systems: Proceedings of the
  7th International Conference on Green Intelligent Transportation System and
  Safety 7}, pages 763--775. Springer, 2018.

\bibitem{douglas1972price}
George~W. Douglas.
\newblock Price regulation and optimal service standards: The taxicab industry.
\newblock {\em Journal of Transport Economics and Policy}, 6(2):116--127, 1972.

\bibitem{beesley1983information}
Michael~E Beesley and Steven Glaister.
\newblock Information for regulating: the case of taxis.
\newblock {\em The economic journal}, 93(371):594--615, 1983.

\bibitem{yang2010nonlinear}
Hai Yang, CS~Fung, KI~Wong, and Sze~Chun Wong.
\newblock Nonlinear pricing of taxi services.
\newblock {\em Transportation Research Part A: Policy and Practice},
  44(5):337--348, 2010.

\bibitem{yang2005regulating}
Hai Yang, Min Ye, Wilson~H Tang, and Sze~Chun Wong.
\newblock Regulating taxi services in the presence of congestion externality.
\newblock {\em Transportation Research Part A: Policy and Practice},
  39(1):17--40, 2005.

\bibitem{camerer1997labor}
Colin Camerer, Linda Babcock, George Loewenstein, and Richard Thaler.
\newblock Labor supply of new york city cabdrivers: One day at a time.
\newblock {\em The Quarterly Journal of Economics}, 112(2):407--441, 1997.

\bibitem{kHoszegi2006model}
Botond K\H{o}szegi and Matthew Rabin.
\newblock {A Model of Reference-Dependent Preferences}.
\newblock {\em The Quarterly Journal of Economics}, 121(4):1133--1165, 11 2006.

\bibitem{KszegiUtility}
Botond K\H{o}szegi.
\newblock Utility from anticipation and personal equilibrium.
\newblock {\em Economic Theory}, 44(3):415--444, 2010.

\bibitem{farber2005tomorrow}
Henry~S Farber.
\newblock Is tomorrow another day? the labor supply of new york city
  cabdrivers.
\newblock {\em Journal of political Economy}, 113(1):46--82, 2005.

\bibitem{doran2014long}
Kirk Doran.
\newblock Are long-term wage elasticities of labor supply more negative than
  short-term ones?
\newblock {\em Economics Letters}, 122(2):208--210, 2014.

\bibitem{solow1956contribution}
Robert~M Solow.
\newblock A contribution to the theory of economic growth.
\newblock {\em The quarterly journal of economics}, 70(1):65--94, 1956.

\bibitem{ashenfelter2010shred}
Orley Ashenfelter, Kirk Doran, and Bruce Schaller.
\newblock A shred of credible evidence on the long-run elasticity of labour
  supply.
\newblock {\em Economica}, 77(308):637--650, 2010.

\bibitem{thakral2017daily}
Neil Thakral and Linh~T T{\^o}.
\newblock Daily labor supply and adaptive reference points.
\newblock Working paper, Harvard University, Department of Economics, 05 2017.

\bibitem{buchholz2016semiparametric}
Nicholas Buchholz, Haiqing Xu, and Matthew Shum.
\newblock Semiparametric estimation of dynamic discrete-choice models.
\newblock {\em arXiv preprint arXiv:1605.08369}, 2016.

\bibitem{wold2004partial}
Herman Wold.
\newblock Estimation of principal components and related models by iterative
  least squares.
\newblock {\em Multivariate analysis}, pages 391--420, 1966.

\bibitem{rashid2015methodological}
Kushairi Rashid and Tan Yigitcanlar.
\newblock A methodological exploration to determine transportation disadvantage
  variables: the partial least square approach.
\newblock {\em World Review of Intermodal Transportation Research},
  5(3):221--239, 2015.

\bibitem{Haenlein2004AB}
Michael Haenlein and Andreas~M Kaplan.
\newblock A beginner's guide to partial least squares analysis.
\newblock {\em Understanding statistics}, 3(4):283--297, 2004.

\end{thebibliography}


\begin{thebibliography}{2}
\providecommand{\natexlab}[1]{#1}

\bibitem[{Romm(2006)}]{Rom06}
Romm, J., \emph{Hell and High Water: The Global Warming Solution}. Harper
  Collins, New York, 2006.

\bibitem[{Guo and Bhat(2007)}]{GuoBha07}
Guo, J.~Y. and C.~R. Bhat, Population Synthesis for Microsimulating Travel
  Behavior. In \emph{Transportation Research Record: Journal of the
  Transportation Research Board, No. 2014}, Transportation Research Board of
  the National Academies, Washington, D.C., 2007, pp. 92--101.

\end{thebibliography}

\end{document}